\documentclass[conf]{new-aiaa}
\usepackage[utf8]{inputenc}
\usepackage{multirow}
\usepackage{graphicx}
\usepackage{amsmath}
\usepackage[version=4]{mhchem}
\usepackage{siunitx}
\usepackage{longtable,tabularx}
\usepackage{booktabs}
\usepackage{xcolor}
\usepackage{colortbl}
\usepackage{pifont}
\usepackage{svg}

\title{Warp Factory: A Numerical Toolkit for the Analysis and Optimization of Warp Drive Geometries}

\author{Christopher Helmerich\footnote{Email: cdh0028@uah.edu} and Jared Fuchs.}
\affil{{The University of Alabama in Huntsville, Huntsville, AL, United States.}}

\author{Alexey Bobrick}
\author{Brandon Melcher}
\author{Luke Sellers}
\author{Gianni Martire}
\affil{Advanced Propulsion Laboratory at Applied Physics, New York, NY, United States}

\begin{document}

\maketitle

\begin{abstract}
The last few decades of warp drive research have focused on analytic methods to explore warp solutions to Einstein’s field equations. These analytic solutions tend to favor simple metric forms which are easier to analyze but limit the space of exploration. In addition, all solutions to date have involved unphysical qualities, such as negative energy, violation of energy conditions, or enormous energy requirements. In an effort to explore the space of physically meaningful warp drives, the Advanced Propulsion Laboratory (APL) at Applied Physics has developed Warp Factory, a toolkit written in MATLAB for numerically analyzing and optimizing warp drive geometries. Warp Factory consists of a series of three primary modules: the solver, the analyzer, and the optimizer. Together, these modules allow users to solve the Einstein field equations, compute energy conditions and scalars, and perturbatively optimize general metrics. Finally, the toolkit offers insightful 2D and 3D visualizations of general metrics and stress-energy tensors. The methods used in Warp Factory, along with their application in evaluating and optimizing common metrics, are discussed. With Warp Factory, APL hopes to accelerate warp research and bring us one step closer to physical and realizable warp drives. 
\end{abstract}

\section{Nomenclature}

{\renewcommand\arraystretch{1.0}
\noindent\begin{longtable*}{@{}l @{\quad=\quad} l@{}}
$\alpha$ & lapse rate \\
$\beta_i$ & shift vector \\
$\gamma_i$ & spatial terms \\
$\kappa_i$ & shear terms \\
$\Gamma^\alpha_{\mu \nu}$ & Christoffel symbol \\
$\rho_E$ & energy density \\
$\rho_{NEC}$ & null energy condition \\
$\nu_{WEC}$ & weak energy condition \\
$k^{a}$ & null vector field \\
$X^{a}$ & timelike vector field \\
$\textbf{X}$ & spacetime 4-vector \\
$p_\sigma$ & momentum density \\
$S_{\alpha \beta}$ & pressure / shear tensor elements \\
$\eta_{\mu \nu}$ & Minkowski metric \\
$M^\mu_\nu$ & frame transformation matrix \\
$g_{\mu \nu}$ & metric tensor \\
$T^{\mu \nu}$  & stress-energy tensor \\
$R_{\mu \nu}$  & Ricci tensor\\
$R$  & Ricci scalar\\
\end{longtable*}}
\clearpage

\section{Introduction}
\lettrine{S}{ince} the publication of Introducing Physical Warp Drives \cite{2021CQGra..38j5009B} the Advanced Propulsion Laboratory (APL) at Applied Physics has continued its exploration of physically meaningful warp drives. To date, the efforts in warp research have been constrained to an analytical approach, which has often provided a limited, and in some cases, inaccurate, assessment of the viability of proposed metrics \cite{2022PhRvD.105f4038S}. To address the challenges of analyzing warp drives, the APL team has developed a numerical analysis toolkit called Warp Factory, which provides a fully generalized solver of the Einstein Field Equations to evaluate the stress-energy tensor along with important scalars and energy conditions. In addition, Warp Factory also provides a novel optimization approach that uses a perturbation-based technique to generate new metrics that can optimize the physicality conditions of a starting warp metric.

In the following paper, an overview of the Warp Factory codebase and its analysis of common warp metrics will be presented. In Section \ref{sec:Background}, the concepts and general equations of General Relativity (GR) are presented. In Section \ref{sec:Toolkit}, the modules of the code are described along with its implementation. In Section \ref{sec:MetricEval}, popular and recently proposed metrics are numerically evaluated and analyzed. In Section \ref{sec:Optimization}, the optimization algorithms are demonstrated. Finally, in Section \ref{sec:Conclusion}, the progress and development of the Warp Factory codebase along with the future work planned for Warp Factory and the APL team will be summarized. 

\section{Background}\label{sec:Background}
\subsection{Metric and Stress-Energy Tensor}
GR is a geometric theory that describes the effects of gravity as a curvature of spacetime. It provides the mathematical machinery to connect the curvature of spacetime with the energy, pressures, and stresses needed to create that curvature. At the center of this theory are the Einstein Field Equations (EFE), which connect the metric tensor $g_{\mu\nu}$ to the stress-energy tensor $T_{\mu\nu}$ via the Ricci curvature $R_{\mu\nu}$ and Ricci scalar $R$:
\begin{equation}
    R_{\mu \nu} - \frac{1}{2}R g_{\mu \nu} = \frac{8 \pi G}{c^4} T_{\mu \nu}
\end{equation}
where the Ricci curvature $R_{\mu \nu}$ is:
\begin{equation}
R_{\mu \nu} =\sum_{\alpha=0}^3 \frac{\partial \Gamma_{\mu \nu}^\alpha}{\partial x^\alpha}-\sum_{\alpha=0}^3 \frac{\partial \Gamma_{\alpha \mu}^\alpha}{\partial x^\nu}+\sum_{\alpha=0}^3 \sum_{\beta=0}^3\left(\Gamma_{\alpha \beta}^\alpha \Gamma_{\mu \nu}^\beta-\Gamma_{\mu \beta}^\alpha \Gamma_{\alpha \nu}^\beta\right)
\end{equation}
which is constructed using the Christoffel symbols $\Gamma_{\alpha \beta}^\gamma$:
\begin{equation}
    \Gamma_{\alpha \beta}^\gamma =\frac{1}{2} \sum_{\sigma=0}^3\left(\frac{\partial g_{\beta \sigma}}{\partial x^\alpha}+\frac{\partial g_{\alpha \sigma}}{\partial x^\beta}-\frac{\partial g_{\alpha \beta}}{\partial x^\sigma}\right) g^{\gamma \sigma} 
\end{equation}
The Ricci Scalar $R$ is:
\begin{equation}
R = \sum_{\mu=0}^3\sum_{\nu=0}^3g^{\mu \nu}R_{\mu \nu}
\end{equation}

Henceforth, all Greek indices will range from 0 to 3, whereas Latin indices range from 1 to 3. Summations will be implied following Einstein notation. 

The metric tensor is a 4x4 matrix:
\begin{equation}
    g_{\mu \nu} = \begin{bmatrix} -\alpha^2 + \beta^i \beta_i & \beta_i \\
    \beta_j & \gamma_{ij} \\
    \end{bmatrix}
\end{equation}
Here the metric terms are expressed following the 3+1 formalism \cite{NumRel_pres} and using the $(-+++)$ metric signature. It should be noted that the metric is symmetric, hence the duplicated indices across the diagonal. In 3+1 the $\alpha$ is the lapse rate which describes the passage of time, $\beta_i$ is the shift vector describing spatial motion, and $\gamma_{ij}$ are the spatial terms which describe spatial shearing ($i \neq j$) and the expansion or contraction of space ($i = j$). Likewise, the stress-energy tensor is also a 4x4 matrix given by:
\begin{equation}
    T_{\mu \nu} = \begin{bmatrix} T_{00} & T_{0i}  \\
    T_{0j} & T_{ij} \\
    \end{bmatrix}
\end{equation}
like the metric tensor, this is also symmetric. The $T_{00}$ represents the energy density, the $T_{0i}$ is the momentum density of the associated energy, $T_{ii}$ are the pressures, and $T_{ij}$ are the shear stresses, as observed by ``at rest'' observers.

When constructing the metric, both the coordinate system used to express the metric and the physical frame are important to consider. They are defined as follows:
\begin{itemize}
\item \textit{Coordinate System:} The spacetime grid selected to represent the space; examples are Cartesian and spherical coordinate systems.
\item \textit{Frame:} In addition to the coordinate system, the frame of observation defines an additional meaning to the coordinates; an example is a comoving frame where time-varying elements are rolled into the definition of the coordinates themselves. Frames define what observers see from their perspective. 
\end{itemize}
For warp metrics, the selection of the coordinate system can reduce the amount of computation by exploiting relevant symmetries. Since warp drives have a direction of travel it is natural to conduct analysis in an axisymmetric coordinate system, which exploits a $\phi$ symmetry in the metric around the direction of travel. Additionally, performing analysis in a comoving frame can reduce the need to consider time derivatives for constant velocity warp drives when solving the EFE.

When interpreting the metric and stress-energy tensors it is important to consider what frame these are defined within, as this will impact the meaning of the terms. Different observers in general will see different values. Frame transformations can be done directly on the stress-energy and metric tensor to transform between observers:
\begin{equation}\label{eq:TensorTransform}
\begin{split}
g_{\mu \nu} = M^{\mu}_{\alpha} M^{\nu}_{\beta} g_{\alpha \beta}\\
        T^{\mu \nu} = M^{\mu}_{\alpha} M^{\nu}_{\beta} T^{\alpha \beta}   
\end{split}
\end{equation}
where $M^{\mu}_{\nu}$ is a transformation matrix. This is defined by a required transformation to shift the metric from one frame into another, which is then also applied to the stress-energy tensor as shown in Equation \ref{eq:TensorTransform}. 

Finally, all observable elements are found by taking the stress-energy tensor and contracting it along a defined observer. The contraction process determines how a given measurable quantity is determined from the tensor as measured by the observer. For any observer, the measurable energy density might involve combinations of different tensor terms when compared from one frame to another. The tensor is not invariant between frames. In general, the contraction for energy density $\rho$, momentum density $p_{i}$ and stresses $S_{ij}$ are given by:
\begin{equation}
    \rho_E = T_{\mu\nu}n^\mu n^\nu
\end{equation}
\begin{equation}
    p_\sigma = -T_{\mu\nu}n^\mu \gamma^\nu_\sigma
\end{equation}
\begin{equation}
    S_{i j} = T_{\mu\nu} \gamma^\mu_i\gamma^\nu_j
\end{equation}
where $n^\mu$ and $\gamma^\mu_\nu$ are the projection vectors of the observer. Other scalar values will be discussed in Section \ref{sec:Scalars}.

\subsection{Warp Structure}
In this analysis of warp drives, some common terminology will be used in describing their geometry and features throughout this paper. These important terms are described below:
\begin{itemize}
\item \textit{Passenger Volume:} The inner volume where the "passengers" or cargo of the warp drive exists, shifted along by the warp bubble around it.

\item \textit{Warp Bubble / Bubble Region:} The non-flat region which transitions the passenger volume to the world boundary and where the required matter-energy of the drive exists.

\item \textit{Observer:} Someone or something moving along a timelike future-directed worldline who can make local measurements to the world around them. These observers are expressed as having 4-vectors with their time and spatial terms restricted to values satisfying a timelike condition\footnote{Timelike vector is defined as a 4-vector, $X$, such that $\langle X,X \rangle = g_{\mu\nu} X^\mu X^\nu < 0$}. A special observer, often used in the analysis of metrics and tensors, is the \textit{Eulerian observer}. This is an observer that resides within a frame locally at rest. In spacetime, this means an observer not moving through space but progressing forward in time.
\end{itemize}
The APL team's working definition of a warp drive is a metric structure that has and/or creates the following effects on observes in spacetime:
\begin{enumerate}
\item \textit{Drags passengers along with the moving warp bubble.} This is a unique feature of a warp metric, as it must provide an effect on observers which is different from having a force applied directly.

\item \textit{Flat or constant spacetime within the passenger volume.} The inner region should be flat according to any and all observers situated within the warp bubble.
\end{enumerate}
There are a few other unique properties that warp drives can provide compared to typical modes of transportation. These are useful and desirable features but not requirements for a warp drive. The features of interest are described below:
\begin{itemize}
    \item \textit{Contactless acceleration of passengers}. Does the warp structure give an effective acceleration to observers without a force imparted to the passengers? This could allow very high accelerations to be achieved without stress on the passengers or cargo inside.
    \item \textit{No momentum flux to infinity}. Does the drive shed its matter-energy to accelerate observers? Likely a change in the energy is required to accelerate the bubble, but whether this is dominated by gravitational waves or classical ejection of matter, like a rocket, is an open topic \cite{2022}.
\end{itemize}

\subsection{Physicality}\label{sec:physicality}
A meaningful warp metric should be something which is physical and hopefully constructable. The physical nature of a warp drive can be expressed as a set of conditions applied to the stress-energy tensor. These conditions are relationships between what any observer would see in the energy density, momentum, pressures, and stresses that align with what would be physical. These energy conditions are the following: 
\begin{itemize}
\item \textit{Non-negative energy in the Eulerian frame}. Energy density should be positive in the locally flat frame of an Eulerian observer.

\item \textit{Satisfies the Null Energy Condition (NEC) everywhere.} The NEC expresses the observed mass-energy density for null (light-ray) observers. For this to be physical, mass-energy density must be non-negative as viewed by any null or lightlike observer.

\item \textit{Satisfies the Weak Energy Condition (WEC) everywhere.} The WEC expresses the observed mass-energy density for timelike (matter) observers. For this to be physical, mass-energy density must be non-negative as viewed by any timelike observer.

\item \textit{Satisfies the Dominant Energy Condition (DEC) everywhere.} The DEC expresses the velocity of observed matter flow. For this to be physical, the matter should not be seen flowing faster than light by any timelike observer.

\item \textit{Satisfies the Strong Energy Condition (SEC) everywhere.} The SEC expresses the tidal effect of gravity acting on observers. For this to be physical, the gravitational tidal effect should be non-negative, i.e., matter always gravitates towards other matter.  



\end{itemize}
In addition to constraints on the matter-energy conditions, to have a realizable warp metric further conditions should be applied based on the practical limits to construct the warp drive:
\begin{itemize}
    \item \textit{Known stress-energy tensor form}. The required elements of the stress-energy tensor should follow those of physically known systems. An example of an unknown stress-energy tensor is one that has pressure terms but no associated energy density.
    \item \textit{Uses reasonable mass}. Any warp bubble should require a practically attainable amount of matter-energy.
\end{itemize}
Finally, while not an immediate requirement for a physical warp drive, it is of interest to consider the properties when the velocity of the system goes superluminal, namely: \textit{are physicality and causality maintained during superluminal motion?} 

In this paper, the focus will be on evaluating a few of the energy conditions, limited to only the Eulerian energy, NEC, and WEC.

\section{Toolkit Structure}\label{sec:Toolkit}
\begin{figure}[h]
\centering
\includegraphics[width=0.8\textwidth]{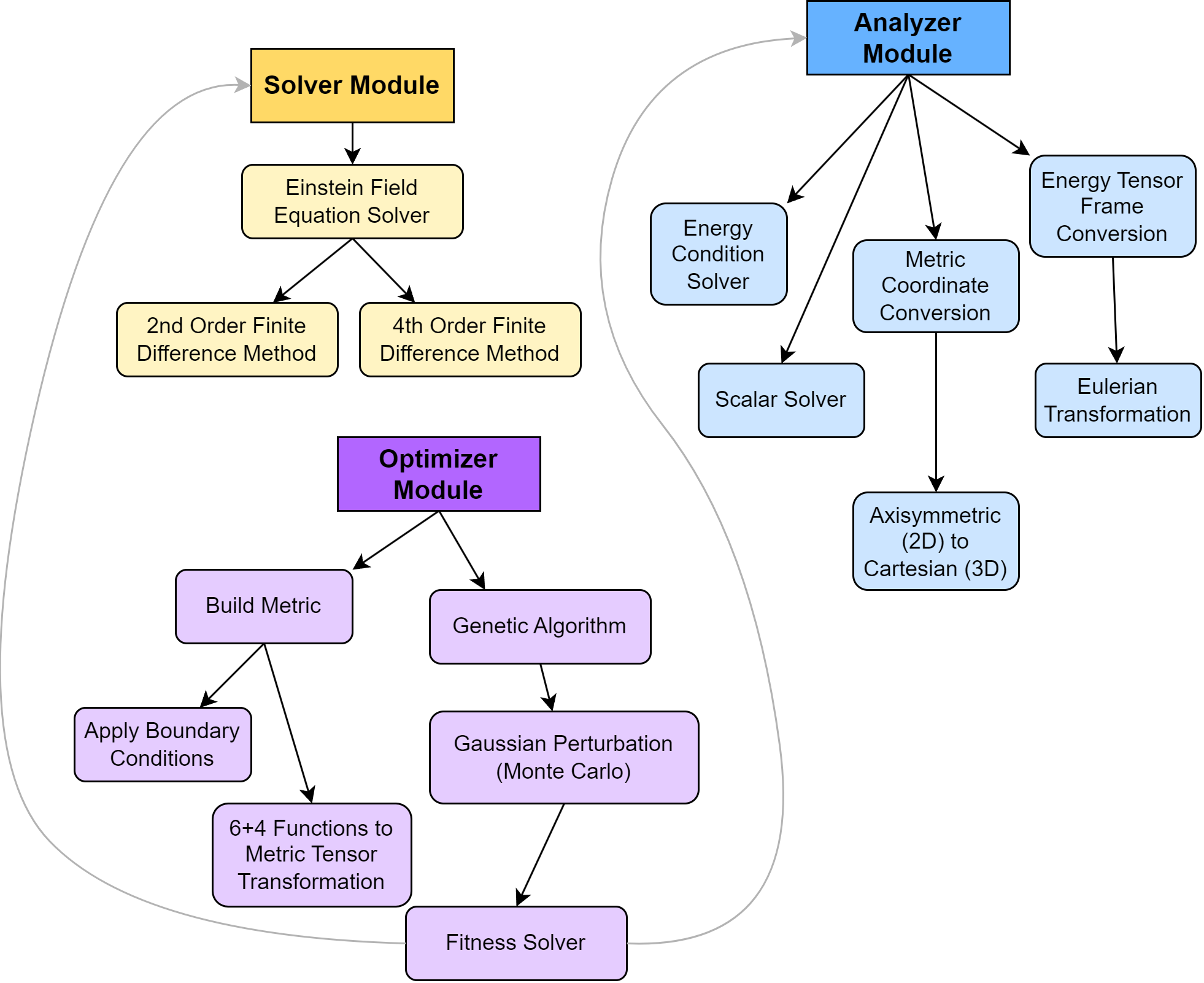}
\caption{Diagram of the major functions and structure of the toolkit. Note that some blocks do reference each other outside of this general hierarchy. Also not shown are smaller functions and applications for plotting data.}
\label{fig:toolkitdiagram}
\end{figure}

The Warp Factory code is built using MATLAB and consists of several functions, organized into modules, that perform different aspects of the overall analysis, visualization, and optimization. In Figure \ref{fig:toolkitdiagram}, a basic structure of the Warp Factory codebase is shown. There are three main modules within Warp Factory: the Solver, Analyzer, and Optimizer.

\subsection{Solver Module}
The solver module handles the solution methods to the EFE. The EFE are solved using a second and fourth-order central finite difference scheme, employed on the spacetime grid \cite{shibata_2016}. The solver function can use either the CPU or GPU of the host machine, with the selection depending on the grid size for best performance. For a basic Cartesian grid, the finite difference solvers assume the derivatives at the bounds are zero. To extend this solver to other coordinate systems, padding points are required for the correct computation of derivatives with radial coordinates or periodic boundaries. At this time, Warp Factory has focused on implementing only the Cartesian and cylindrical coordinate systems. For a 2D grid in cylindrical or axisymmetric coordinate frames, the addition of "ghost points" are required for the computation of derivatives around $r = 0$. These ghost points mirror the real points across $r$ so that the minimum grid points are provided for the finite difference method at $r=0$. The mirroring process also depends on the metric component, with some terms being mirrored symmetrically, whereas other terms are mirrored anti-symmetrically. If the size of the spacetime is defined to be unity along any dimension (for example, the time or a symmetric direction like $\phi$), the solver assumes the derivatives along those dimensions are zero.

\subsection{Analyzer Module}
The analyzer module takes in the metric and stress-energy tensor and generates observables for the user. This module also conducts the relevant transformations and contractions to generate these values. Two kinds of transformations can be performed: frame or coordinate transformations.

\subsubsection{Frame Transformations}
Warp Factory currently implements a frame transformation solver to convert the stress-energy tensor into an Eulerian frame from any arbitrary input metric. The Eulerian frame is defined as the frame which has a flat Minkowski metric:
\begin{equation}
    \eta_{\mu\nu} = \text{diag}(-1,x_1,x_2,x_3)
\end{equation}
The coordinate system impacts the frame transformation\footnote{Note that typically the Minkowski metric is the same in covariant vs contravarint forms, but this is only the case in Cartesian coordinates and not the case for a general coordinate system.}. For Cartesian, $x = [1, 1, 1]$, but for cylindrical $x = [1, r^2, 1]$. The transformation matrix solves, at all points in spacetime, the matrix equation:
\begin{equation}
    \textbf{M}^T\textbf{g}\textbf{M}=\boldsymbol{\eta}
\end{equation}
where $\textbf{M}$ is a lower triangular matrix. This transformation has a general symbolic solution, albeit a complicated one, which is computed for all points. With $\textbf{M}$ defined, the stress-energy tensor is transformed using Equation \ref{eq:TensorTransform}. The motivation for this transformation is so that further point-wise analysis on the energy tensor components can be conducted within a well-defined frame, such as when determining energy condition violations which require defining a vector field. 

In addition to Eulerian transformation, shifting the metric into the comoving frame saves computation for constant velocity warp drives by removing the need for derivatives along $t$. The transformation acts on the shift vector:
\begin{equation}
    \beta^\prime = \beta - v
\end{equation}
This results in shifting the frame of analysis to that of a frame comoving with the passenger volume and is typically centered in the world volume, with the world boundary moving at $-v$. In the comoving frame, there are no coordinates of the warp drive which need to vary with time, thus all $\partial_t$ terms are removed.

\subsubsection{Coordinate Transformations}
Transformations between the 2D and 3D coordinates can be performed directly on the metric components, mapping the metric tensor from one coordinate to another. This mapping solves the transformations of the line element using standard coordinate relations. When mapping from axisymmetric 2D to Cartesian 3D, points beyond the radius boundary of the original 2D system are set to a specified boundary condition, due to a lack of data to extrapolate. To improve the accuracy of the translation, a 3rd-order Lagrange polynomial interpolation is used. These transformations also are conducted on the vector fields as needed and in the definition of the Minkowski metric, whose diagonal elements change depending on the coordinate system.

Contravariant and covariant transformations are done using the metric tensor acting on vectors or tensors as an index-raising, lowering, or mixing operation:
\begin{equation}
    U_\nu = g_{\mu\nu} U^\mu, \ \ U^\nu = g^{\mu\nu} U_\mu
\end{equation}
\begin{equation}
    T_{\mu\nu} = g_{\nu\beta} g_{\mu\alpha} T^{\alpha \beta},  \ \ T^{\mu\nu} = g^{\nu\beta} g^{\mu\alpha} T_{\alpha \beta} 
\end{equation}
\begin{equation}
    T^\mu_\nu = g_{\nu\alpha} T^{\mu \alpha}
\end{equation}
The metric tensor $g_{\mu\nu}$ is transformed into covariant or contravariant forms using matrix inversion at each point.

\subsubsection{Energy Conditions}
The primary mathematical expression of physicality to a warp metric is in the satisfaction of energy conditions. There are four main conditions used in the literature: null, weak, dominate, and strong \cite{2014arXiv1405.0403C}. Currently, Warp Factory only implements analysis on the null and weak, which are described below. 

The Null Energy Condition (NEC) is expressed at a given point in spacetime as: 
\begin{equation}
    \nu_{NEC}(\textbf{X}) = T_{ab}(\textbf{X}) k^a k^b \ge 0
\end{equation}
Where $T_{ab}$ is the energy tensor transformed to an orthonormal frame at each point in spacetime $\textbf{X}$ and for a future-pointing null vector field $\textbf{k}$ defined (in flat space) as:
\begin{equation}
    \textbf{k} = (t,\textbf{r})
\end{equation}
where the null vector field satisfies the condition of:
\begin{equation}
     \langle \textbf{k}, \textbf{ k} \rangle = 0 \footnote{In GR the inner product is generalized as: $\langle a , b \rangle = g_{\mu\nu} a^\mu b^\nu$ , which differs from 3-vector inner products.}
\end{equation}
The value of $t$ is fixed to unity and all spatial vectors in the field satisfy the condition of $|\textbf{r}| = 1$ for a specified number of vectors. The density of vectors will define the resolution of finding the maximum null violation amount. Each spatial vector is sampled by generating a grid of uniformly distributed points on a sphere of unity radius. This process is done for each spacetime point with a map of the maximum violation returned. While this is often considered as a binary check on the condition of $\nu_{NEC}(\textbf{X}) \ge 0$ across all spacetime, the magnitude of violation can be used as a fitness parameter itself during optimization.

The Weak Energy Condition (WEC) is calculated in a similar matter, but uses a timelike vector field:
\begin{equation}
    \rho_{WEC}(\textbf{X}) = T_{ab}(\textbf{X}) X^a X^b \ge 0
\end{equation}
where \textbf{X} is defined similarly to k but as a timelike vector field satisfying the condition of:
\begin{equation}
    \langle \textbf{X}, \textbf{X} \rangle < 0
\end{equation}
the spatial terms of $\textbf{r}$ in the vector field $\textbf{X}$ are generated evenly across the surface of a sphere as before, but also in a series of shells with decreasing magnitudes of $\textbf{r}$ to 0. These shells are evaluated and the maximum violation is returned for all directions and all shells. It is also important to note that NEC violation implies WEC violation. 

\subsubsection{Scalars}\label{sec:Scalars}
The properties of the metric can be explored by comparing coordinate invariant quantities, specifically the expansion and shear scalars. Each of these scalars is built from the metric tensor and evaluated at each point in spacetime using projections of the metric on a timelike vector field $U^\mu$, defined by the lapse and shift vectors as\footnote{This is just one example of a timelike vector field of interest, these quantities can be computed for any timelike vector field.}:
\begin{equation}
    U^\mu = \frac{1}{\alpha}\left(1,-\beta_i \right) \ \ \ U_\nu = g_{\mu\nu} U^\mu
\end{equation}

The two scalars of interest are expansion and shear. 

The expansion scalar $\theta$ is defined as:
\begin{equation}
   \theta = g^{\mu\nu}\theta_{\mu\nu}
\end{equation}
where $\theta_{\mu\nu}$ is the stress tensor found from the projection tensor $P^\alpha_\mu$ acting on the observer vector field $U$ as
\begin{equation}
    \theta_{\mu\nu} = P^\alpha_\mu P^\beta_\nu \nabla_{(\alpha} U_{\beta)}
\end{equation}
and the projection tensor is given by:
\begin{equation}
    P_{\mu\nu} = g_{\mu\nu} + U_\mu U_\nu
\end{equation}

The shear scalar $\sigma^2$ is defined as:
\begin{equation}
    \sigma^2 = \frac{1}{2} \sigma_{\mu\nu} \sigma^{\mu\nu}
\end{equation}
where $\sigma_{\mu\nu}$ is the shear tensor found from the stress tensor and projection tensor $P_{\mu\nu}$, shear tensor $\theta_{\mu\nu}$ and shear scalar $\theta$ as:
\begin{equation}
   \sigma_{\mu\nu} = \theta_{\mu\nu} - \frac{\theta}{3} P_{\mu\nu}
\end{equation}

These two scalars can be interpreted as invariant quantities signifying certain effects the metric tensor has on the spacetime for the observer $U$. Specifically, they define the following: 
\begin{itemize}
\item \textit{Expansion Scalar:} The expansion scalar can be thought of as a quantity that describes the change in volume of an object. 


\item \textit{Shear Scalar:} The shear scalar can be thought of as a quantity that describes the relative expansion or contraction along any dimensions while preserving the volume of an object.
\end{itemize}

The most common scalar discussed in the literature is expansion. For the Alcubierre metric, this exhibits a contraction in front of the bubble and an expansion behind it. However, it should be noted that the presence of expansion, or any of these scalars being non-zero, is not fundamentally required to make a warp drive \cite{2002CQGra..19.1157N}.

\subsubsection{Momentum Flow}
Most warp metrics generate terms beyond just the $T_{00}$, including pressures, shears, and momentum. The momentum describes the motion of the energy associated with the warp metric. To understand the flow of matter in the bubble a map of flowlines is constructed.  This is done by building a velocity vector field comprised of the momentum flux terms $p_i$ across each spacetime point:
\begin{equation}
    \boldsymbol{\Omega}(\textbf{X}) = \left(p_1(\textbf{X}), p_2(\textbf{X}), p_3(\textbf{X})\right)
\end{equation}
where $T^{\mu \nu}$ is first transformed into the Eulerian frame and contracted to build $p_i$. The flowline is generated using a "virtual" timestep from a starting grid of points, which are pushed along by a velocity imparted by the momentum vector field. Note that this is not a trajectory of matter moving along a geodesic, but rather a visualization that captures the snapshot of the momentum density structure in the warp bubble as generated by the metric tensor. The virtual time steps are not real-time evaluations of the dynamical spacetime across $t$. Momentum flowlines are currently only generated for metrics in Cartesian coordinates, with coordinate transformations performed as needed. Since this flowline is computed on a finite grid, a trillinear interpolation is used on the momentum vector field data for interpolation between points.

\subsection{Optimization Module}\label{sec:OptModule}
The complexity of solving the EFE presents a limitation to any analytical approach in exploring the space of complicated warp metrics. One method to dynamically generate new metrics is through a simple Machine Learning (ML) approach. This technique will perform perturbations to the metric, compute their impact on the stress-energy tensor, and then calculate the resulting energy conditions as a fitness measure on improving the physicality of a warp drive. Each perturbation is selected based on its improvement to the fitness. 

The ML approach is implemented as a Monte Carlo Gaussian Perturbation within a Genetic Algorithm. To allow for general perturbations that will not break the metric, a new formalism is developed called 6+4. Under 6+4, the usual 3+1 approach is extended with modifications in order to preserve the metric signature $(-+++)$ under arbitrary perturbations to the underlying functions which generate it. The form of this metric in terms of the 6+4 functions is:
\begin{equation}
   g_{\mu\nu} = \begin{bmatrix}
    g_{00} & \beta_1 & \beta_2 & \beta_3 \\
   \beta_1&  g_{11} & \kappa_1 & \kappa_2 \\
   \beta_2 & \kappa_1 &  g_{22} & \kappa_3 \\
   \beta_3 & \kappa_2 & \kappa_3 &  g_{33} \\
    \end{bmatrix}
\end{equation}
where each of the diagonal terms is built from the ten independent 6+4 functions $\beta_i, \kappa_i, \gamma_i,$ and $\alpha$ as:
\begin{equation}
    g_{33} = \gamma_{3}^2
\end{equation}

\begin{equation}
    g_{22} = \gamma_{2}^2+\left(\frac{\kappa_3}{\gamma_3}\right)^2
\end{equation}
\begin{equation}
    g_{11} = \gamma_1^2 + \left(\frac{\kappa_2}{\gamma_3}\right)^2 + \left(\frac{\kappa_1 -\frac{\kappa_2\kappa_3}{\gamma_3^2}}{\gamma_2}\right)^2 
\end{equation}
\begin{equation}
    g_{00} = - \alpha^2 + \left(\frac{\beta_3}{\gamma_3}\right)^2 + \left(\frac{\beta_2 - \frac{\beta_3\kappa_3}{\gamma_3^2}}{\gamma_2}\right)^2 + \left(\frac{- \beta_1 + \frac{\beta_3\kappa_2}{\gamma_3^2}  + \frac{(- \beta_2\gamma_3^2 + \beta_3\kappa_3)(- \kappa_1\gamma_3^2 + \kappa_2\kappa_3)}{\gamma_2^2\gamma_3^4}}{\gamma_1}\right)^2 
\end{equation}
During the optimization, changes are made to each of the 6+4 functions and the full metric is then constructed using the equations above. The modifications made to enable 6+4 to be robust in the face of perturbations do not share the same physical interpretations which drive the formalism of 3+1 for a generic metric. While this is important for the perturbation code, the use of 3+1 decomposition in interpreting results should still be used.

\subsubsection{Monte Carlo Gaussian Perturbation Method}
The core element of the ML approach is the Monte Carlo Gaussian Perturbation (MCGP) technique. In MCGP each perturbation is applied, centered on points selected in the spacetime, for each function in the 6+4 formalism. These perturbations are Gaussian functions that can vary in both amplitude and $\sigma$. A minimum $\sigma$ size bound is used to avoid sharp changes, which can exploit issues with the numerical grid resolution in the finite difference method. This process is inherently time-consuming as the perturbations require computation of the EFE and evaluation of energy conditions to determine their impact on fitness. This is improved by doing perturbations to parts of the space and applying symmetry where possible. The selection of which points to perturb are also selectively chosen and currently use the NEC violation map to focus changes around regions that need the most adjustment. The method uses a probability of selection which scales the violation logarithmically, where:
\begin{equation}
    p_{point}(\textbf{X}) = a \ e^{log10(\nu_{NEC}(\textbf{X}))}
\end{equation}
The amplitude of the perturbation is expressed as a fraction of the local metric value to enforce perturbations that are scaled appropriately. 

The usage of symmetry starts by running the MCGP only on 2D axisymmetric grids. In addition to using a symmetric coordinate, perturbations in the $\rho, z$ space are also mirrored around the midpoint along the z-axis (direction of travel), with most terms mirrored symmetrically, and the $\beta_2$, $\kappa_1$, and $\kappa_3$ mirrored anti-symmetrically. The MCGP cycles through making perturbations to each metric function. Which metric functions are allowed to vary are user-selectable. 

The acceptance or rejection of perturbations is based on improvement of fitness. To decrease the chance of getting stuck at a local minimum, the acceptance of the perturbations allows for a certain decrease in the fitness to be accepted, within a user-selectable margin. This tends to allow better improvement in longer runs. The boundaries of the world and passenger volume are held fixed during the MCGP. 

\subsubsection{Genetic Algorithm}
\begin{figure}[h]
\centering
\includegraphics[width=0.85\textwidth]{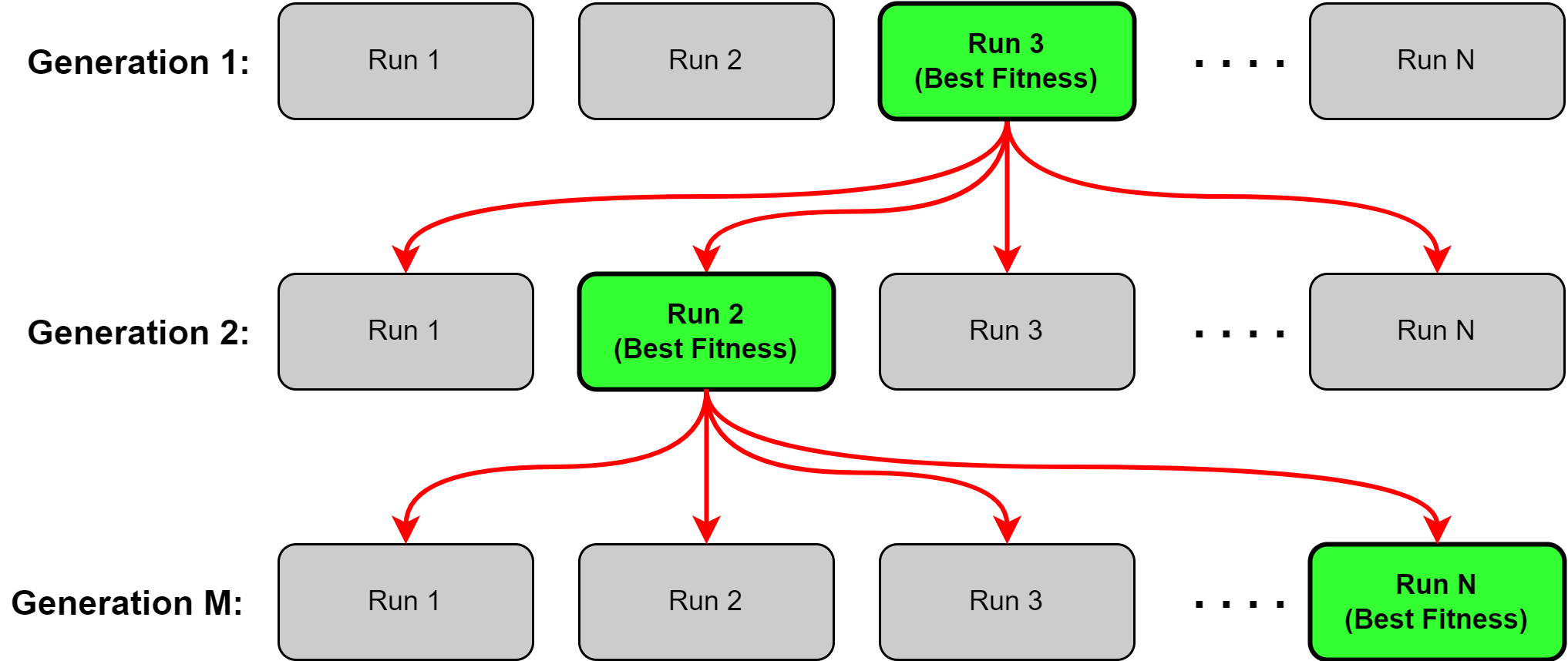}
\caption{Selection and mutation flow using the genetic optimization algorithm.}
\label{fig:GeneticFlow}
\end{figure}

A key feature of the MCGP is the randomness in its perturbations. To improve both the effectiveness and speed of execution, a genetic algorithm is used. This algorithm runs the MCGP on several mutated metrics in parallel. A mutated metric is one that has random changes applied to it. From these parallel runs or populations, the best result can then be selected and further mutated again and run in parallel across several generations. The process flow for this is shown at a high level in Figure \ref{fig:GeneticFlow}.
The mutation process adds Gaussian perturbations to a starting metric regardless of whether it improves or worsens the fitness. The addition of starting mutations allows even more robustness in the face of local minima.

\clearpage
\section{Metric Evaluations}\label{sec:MetricEval}
Using Warp Factory, common and new warp metrics discussed in the literature can be evaluated from a numerical perspective, showcasing them in greater detail.

\subsection{Alcubierre Metric (AM)}
The most famous and very first warp metric was proposed by Alcubierre in 1994 \cite{1994CQGra..11L..73A}. This metric uses a flat passenger volume with a constant shift vector which connects to the boundary by a hyperbolic tangent function. The AM is shifted into a comoving frame, which simply inverts the shift vector to be +v at the world boundary and zero inside the passenger volume; in this frame time derivatives can be ignored. 

\begin{figure}[hbt!]
\centering
\includegraphics[width=\textwidth]{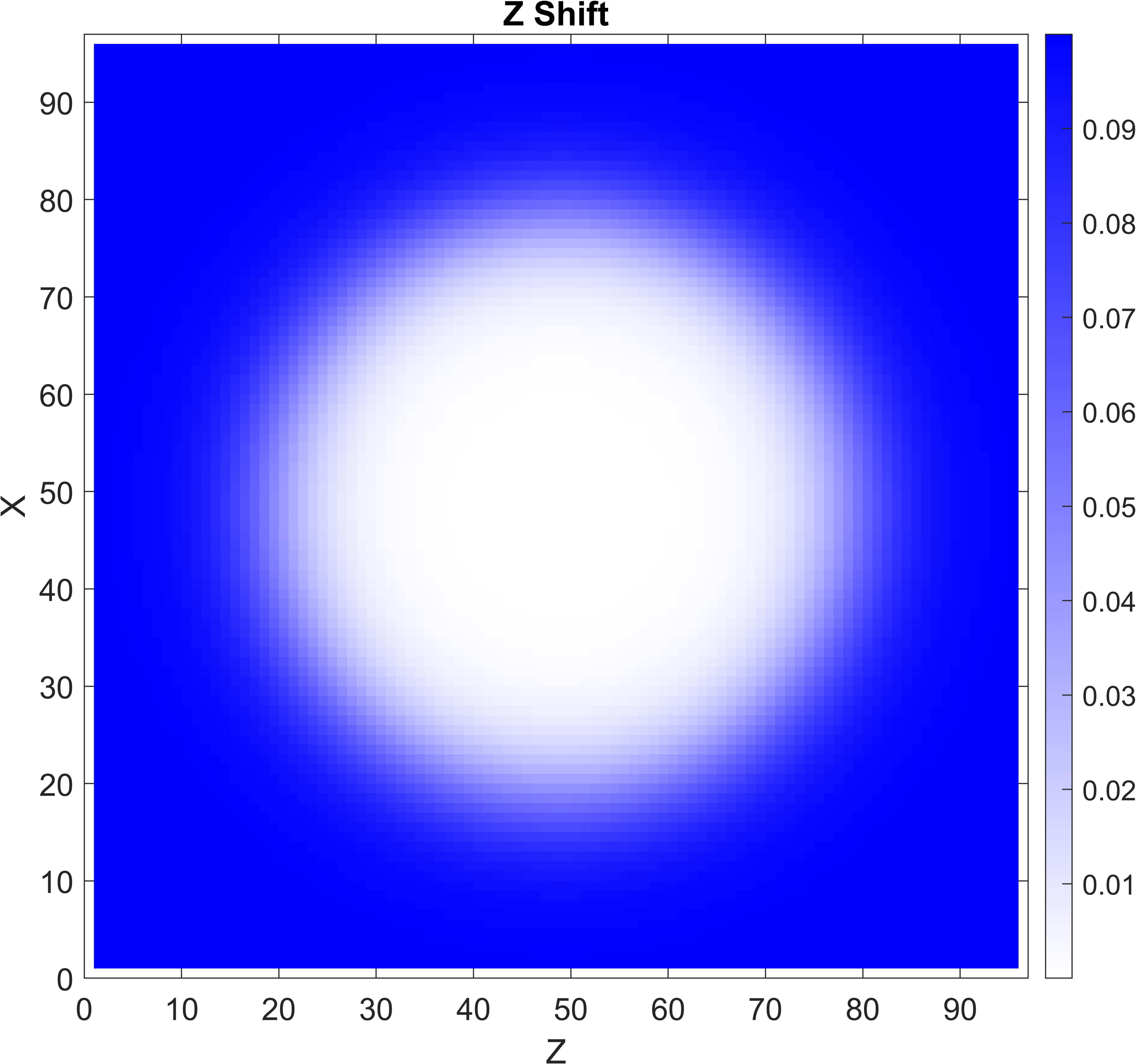}
\caption{Metric functions for the AM. Only the shift vector along Z is modified from Minkowski. Direction of motion is along +Z. Cross-section is plotted for y = 0.}\label{fig:AM_functions}
\end{figure}

\begin{figure}[hbt!]
\centering
\includegraphics[width=\textwidth]{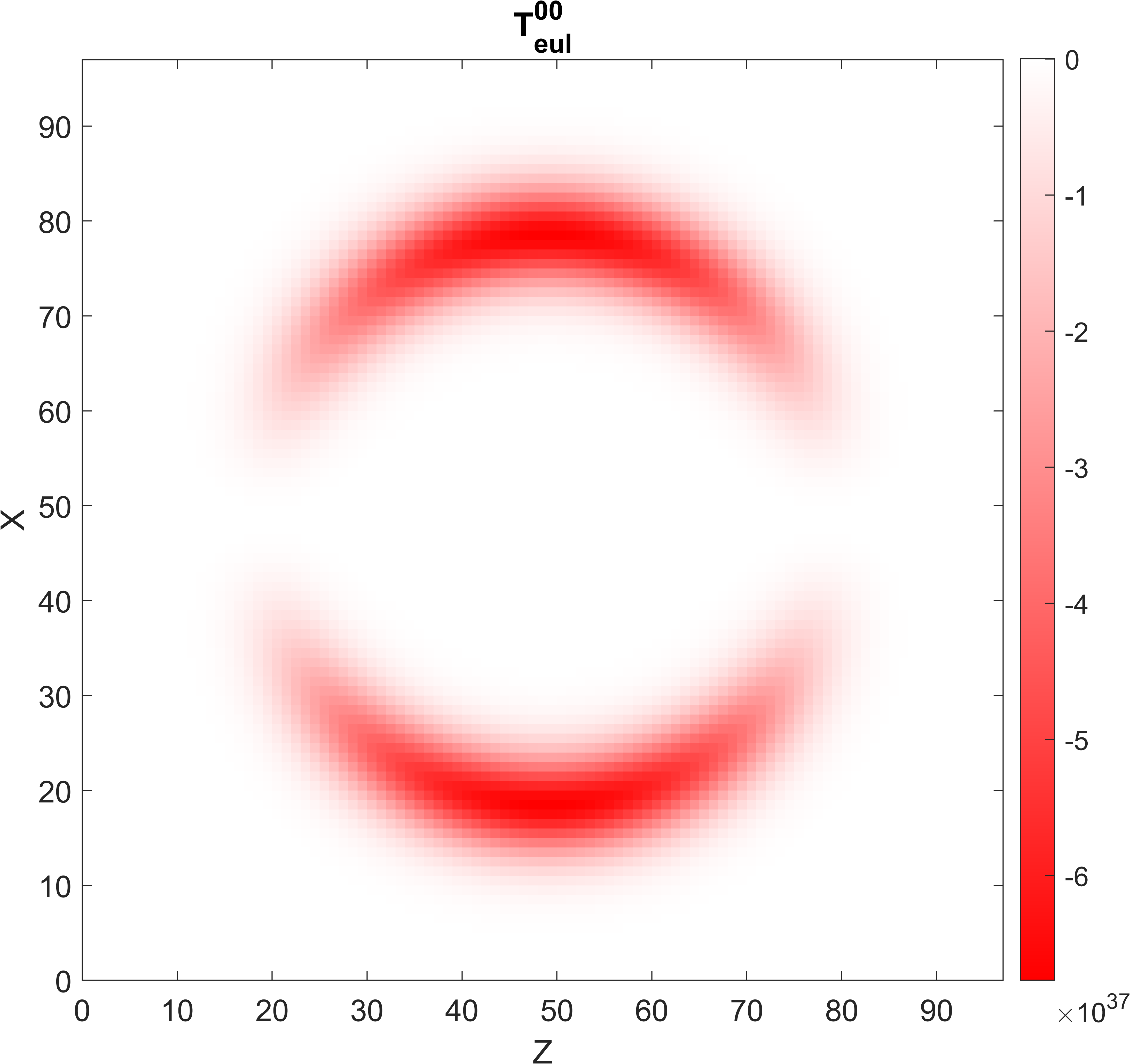}
\caption{Eulerian energy density for the AM. Direction of motion is along +Z. Cross-section is plotted for y = 0.}\label{fig:AM_Eden}
\end{figure}

\begin{figure}[hbt!]
\centering
\includegraphics[width=\textwidth]{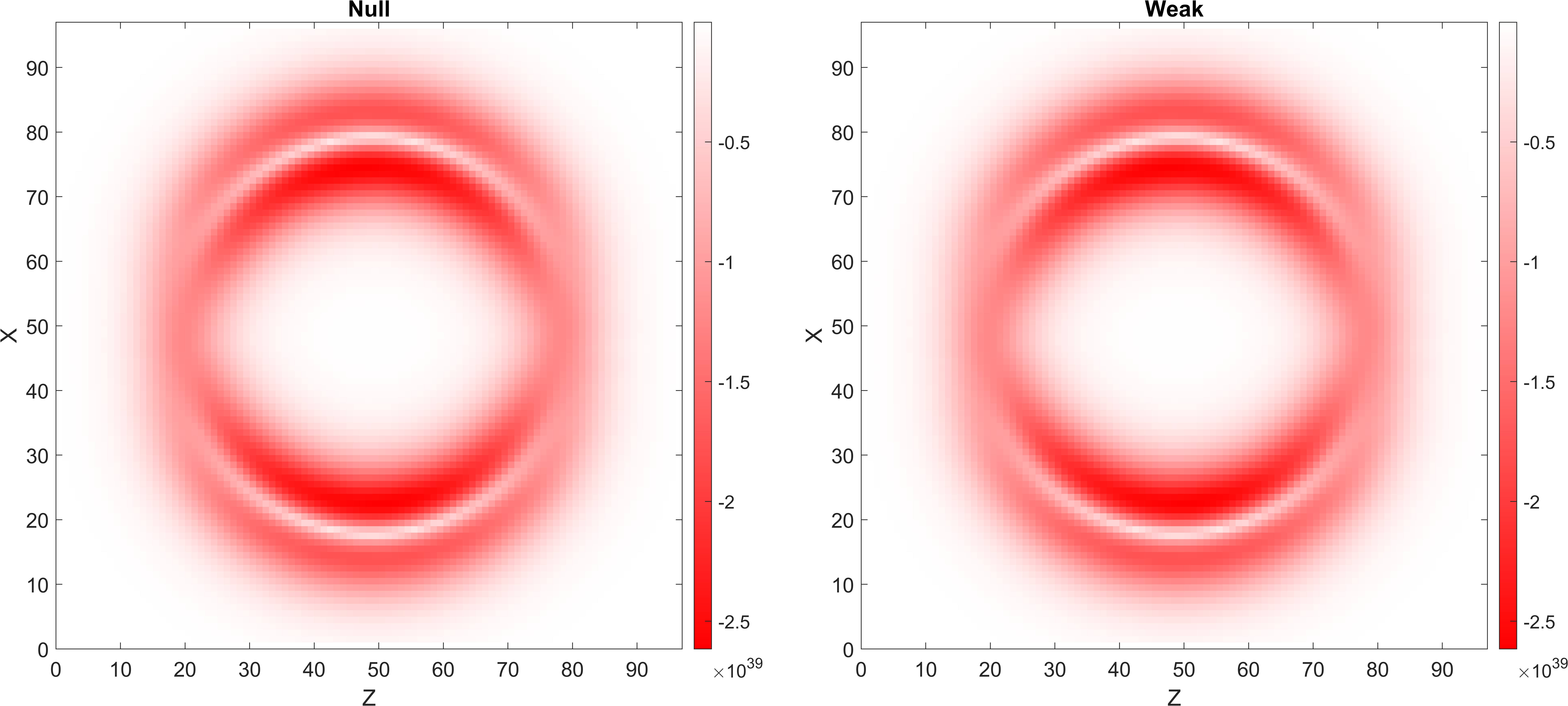}
\caption{NEC and WEC for the AM. Direction of motion is along +Z. Plotted are the most violating values across all observers. Cross-section is plotted for y = 0.}\label{fig:AM_Econd}
\end{figure}

\begin{figure}[hbt!]
\centering
\includegraphics[width=\textwidth]{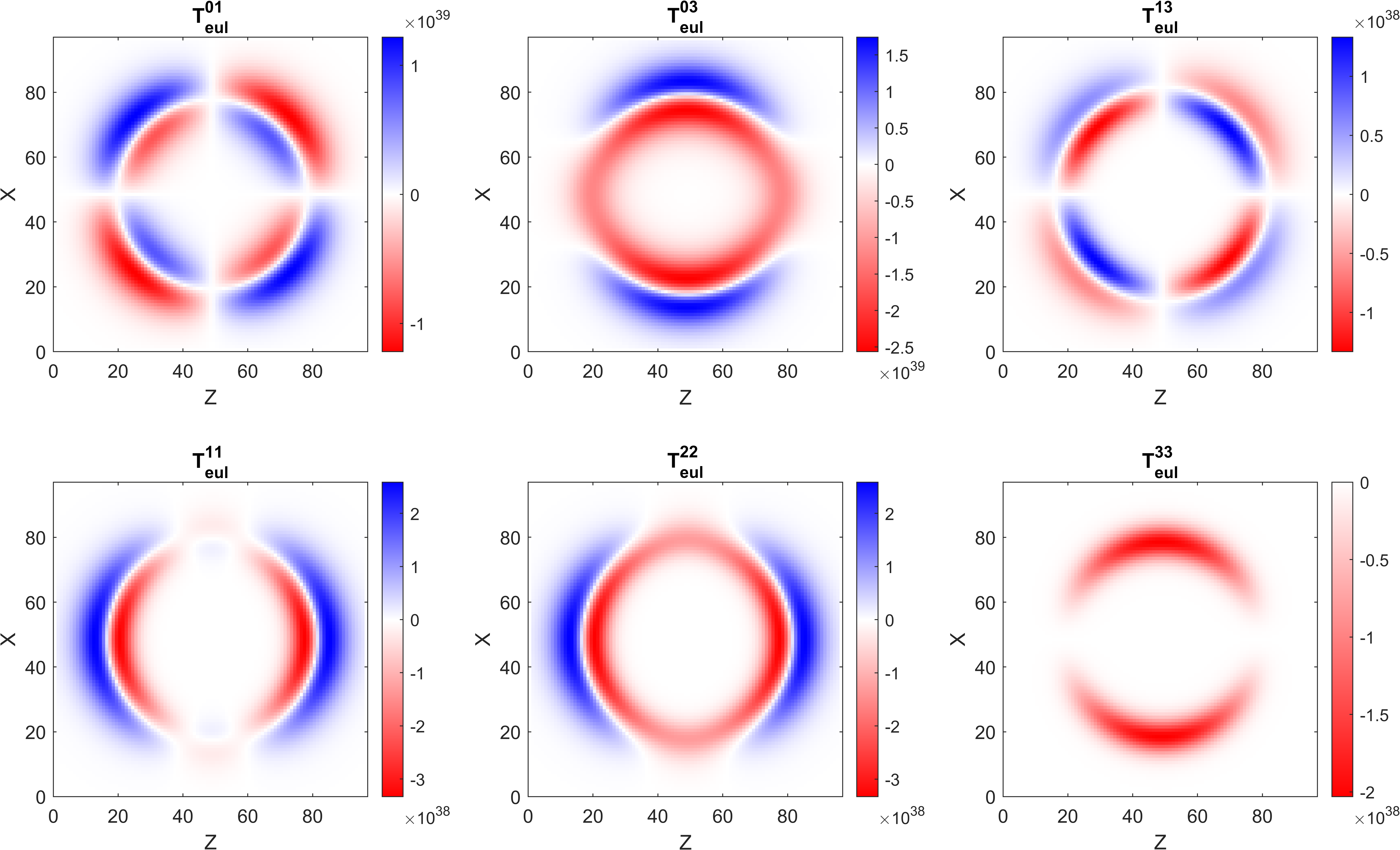}
\caption{The stress-energy tensor, in the Eulerian frame, for the AM. The $T^{00}$ component is shown in Figure \ref{fig:AM_Eden}.  Direction of motion is along +Z. Cross-section is plotted for y = 0.}\label{fig:AM_Etensor}
\end{figure}

\begin{figure}[hbt!]
\centering
\includegraphics[width=\textwidth]{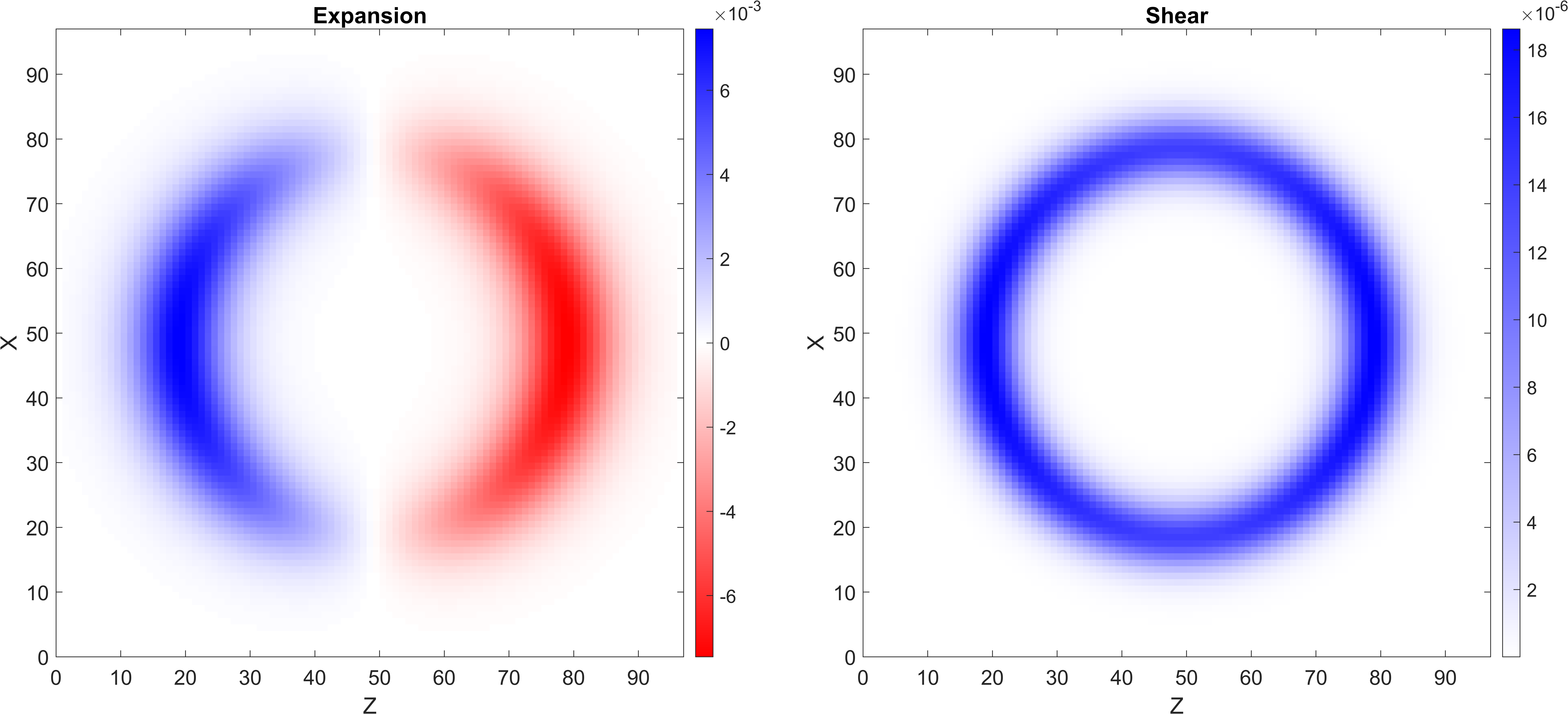}
\caption{Expansion and shear scalars for the AM. Direction of motion is along +Z. Cross-section is plotted for y = 0.}\label{fig:AM_scalars}
\end{figure}

Using Warp Factory, we construct the AM in 3D coordinates and solve the EFE. The shift vector is shown in Figure \ref{fig:AM_functions}, the Eulerian energy density in Figure \ref{fig:AM_Eden}, the NEC and WEC in Figure \ref{fig:AM_Econd}, the full stress-energy components in Figure \ref{fig:AM_Etensor}, and finally the scalars in Figure \ref{fig:AM_scalars}.

As identified by Alcubierre himself, the AM metric violates both the NEC and WEC within the bubble region due to its requirement of negative energy density; this result is confirmed here using the code. 


More generally, the AM metric falls into what is commonly referenced as a Nat\'ario class warp metric \cite{2002CQGra..19.1157N} which is built only using shift vector components that satisfy the form of:
\begin{equation}
  g_{\mu \nu} = \begin{bmatrix}
-\left(1-v^2\right) & -v_i \\
-v_j & \delta_{i j}
\end{bmatrix}
\end{equation}
where $v$ follows the passenger volume of choice. 

\begin{figure}[hbt!]
\centering
\includegraphics[width=\textwidth]{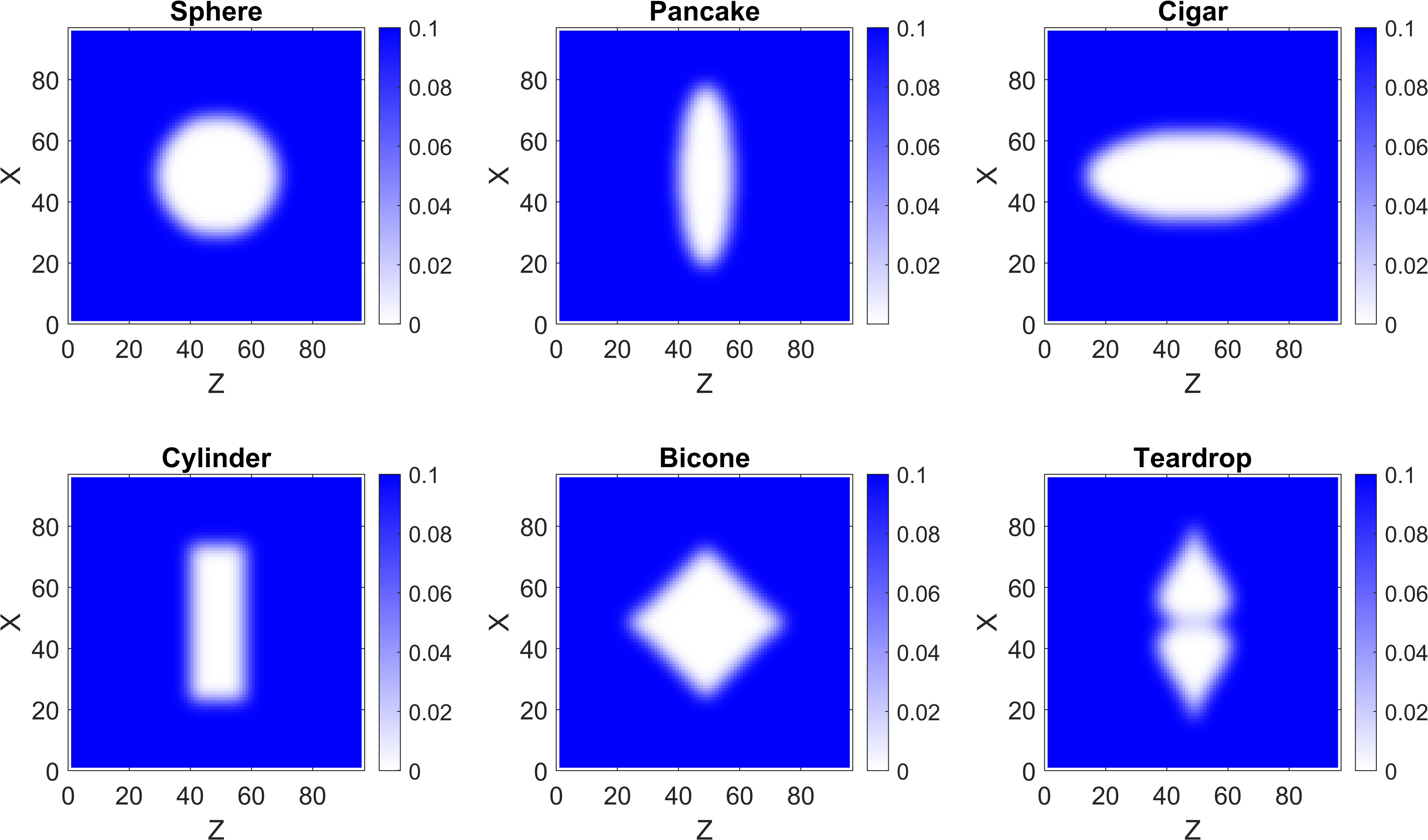}
\caption{Shift vector along the Z axis for various passenger volume shapes. They are defined in a comoving frame. Cross-section is plotted for y = 0.}\label{fig:GenPass_shift}
\end{figure}

\begin{figure}[hbt!]
\centering
\includegraphics[width=\textwidth]{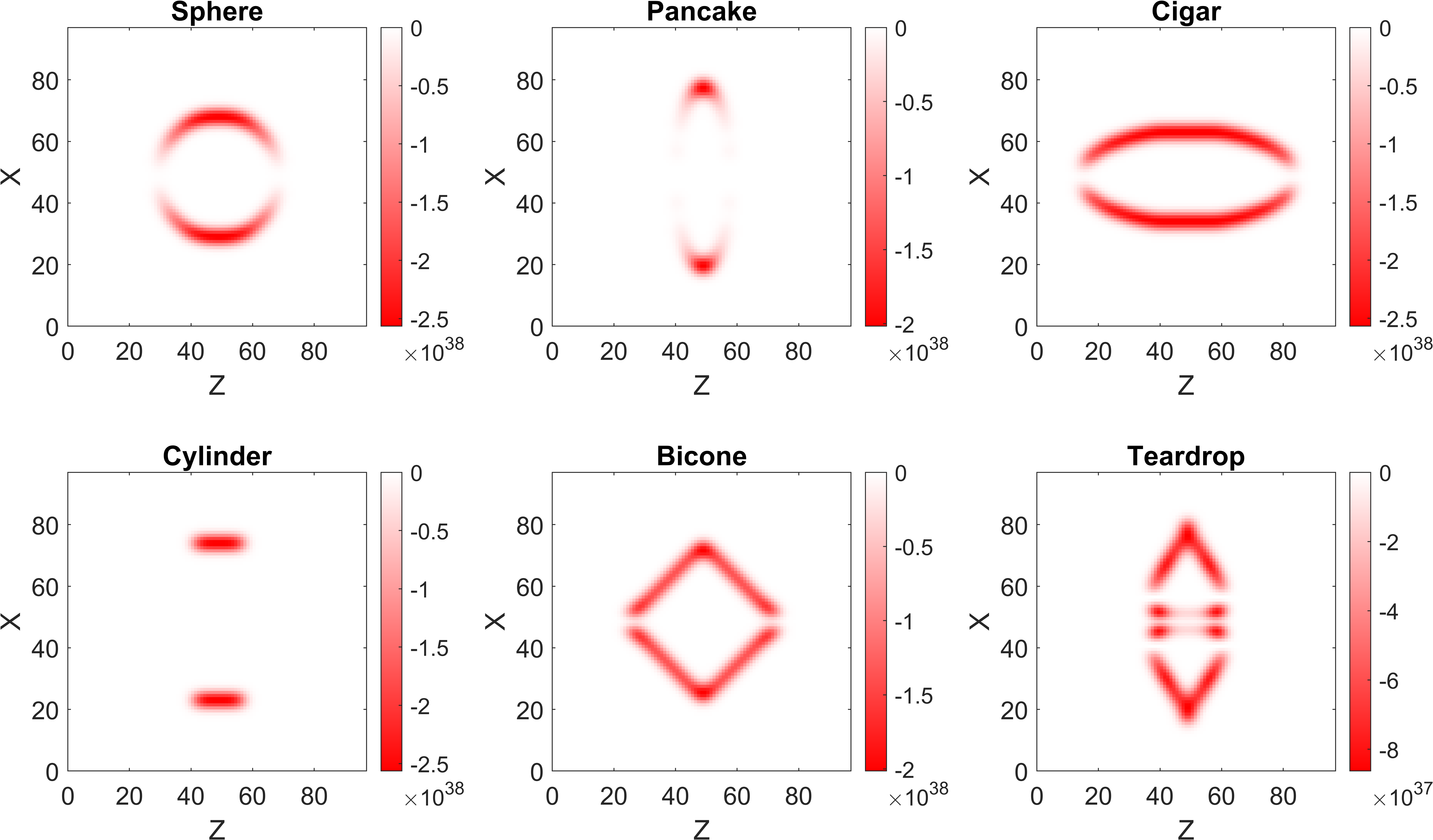}
\caption{Eulerian energy density for each of the passenger volumes. Cross-section is plotted for y = 0.}\label{fig:GenPass_Eden}
\end{figure}

\begin{figure}[hbt!]
\centering
\includegraphics[width=\textwidth]{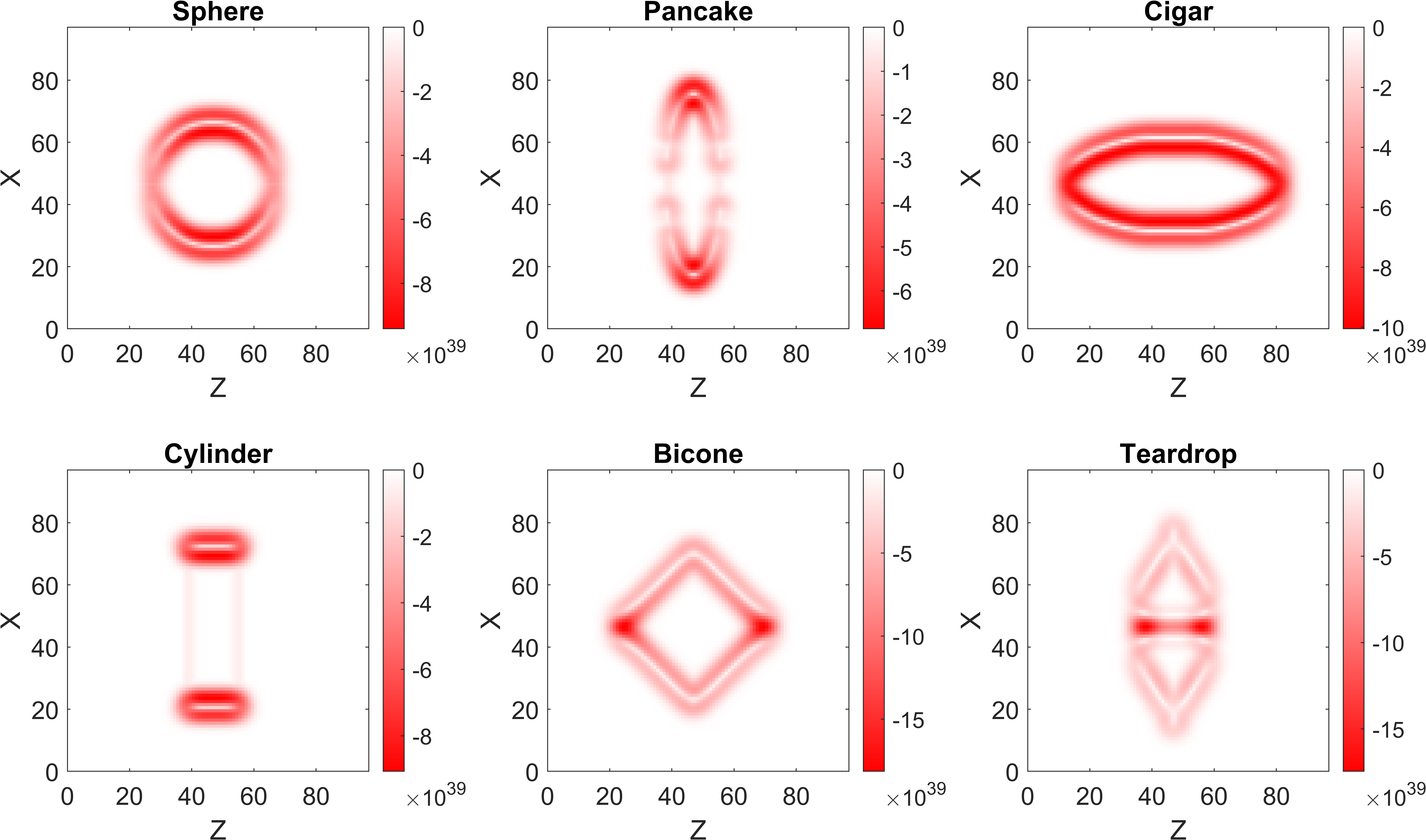}
\caption{NEC for each of the passenger volumes. Cross-section is plotted for y = 0.}\label{fig:GenPass_NEC}
\end{figure}

\begin{figure}[hbt!]
\centering
\includegraphics[width=\textwidth]{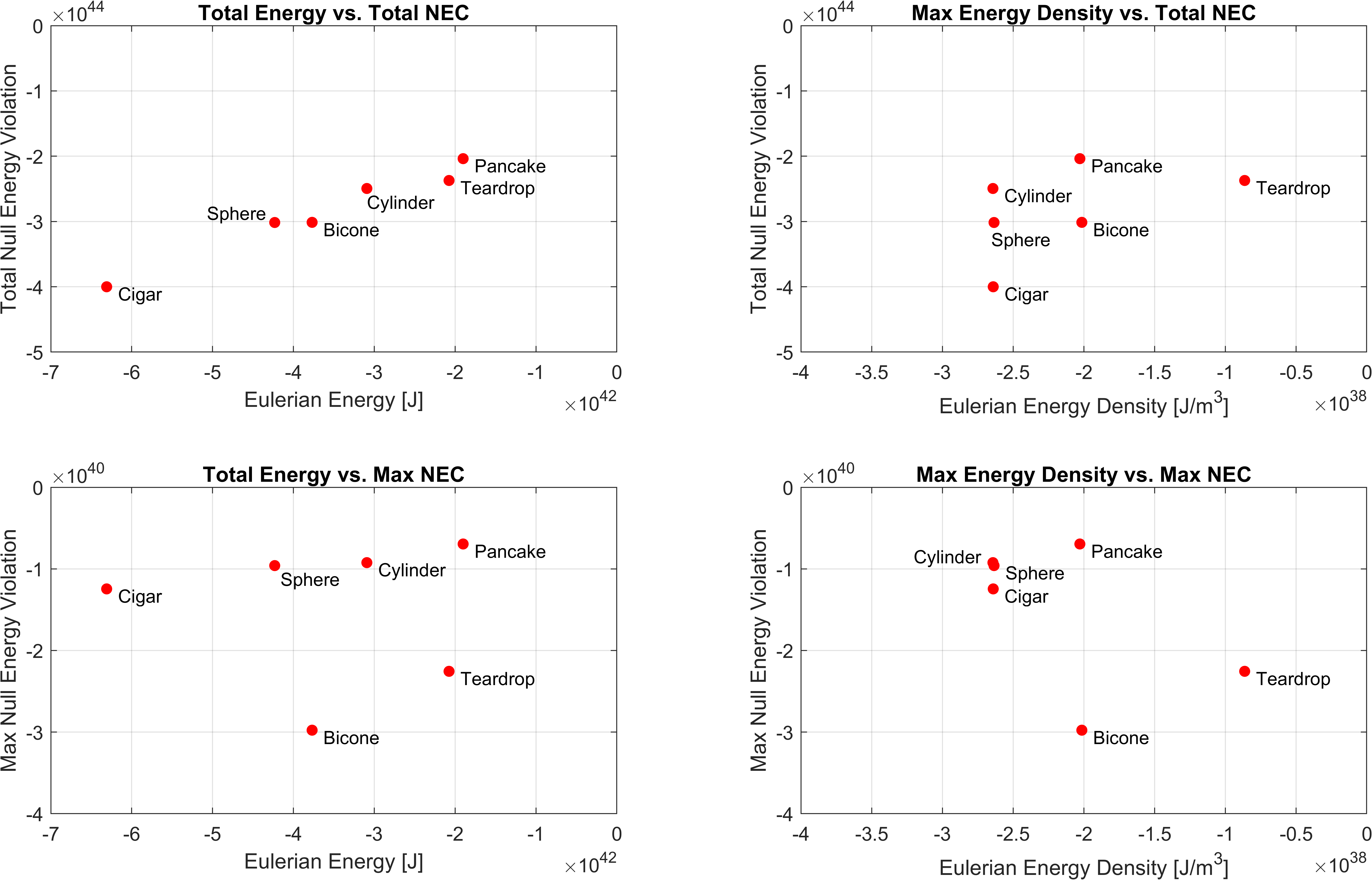}
\caption{Scatter plots of the Eulerian energy density, Eulerian total energy, total NEC violation, and NEC violation density relationships for the various passenger volumes.}\label{fig:GenPass_Scatter}
\end{figure}

With Warp Factory, different and more complicated passenger volumes can be explored to see how the topology of the inner volume impacts the energy conditions. For this analysis, the passenger volume shift velocity is set along Z and is smoothly connected to the boundary using a Gaussian smoothed falloff over a specified radius $R$. The Eulerian energy and NEC are compared between a pancake, cigar, cylinder, bicone, and teardrop passenger volume shapes to that of a standard spherical volume. All shapes maintain a constant total passenger volume for an equal comparison. The shift vectors for each shape are shown in Figure \ref{fig:GenPass_shift}.
The energy density and NEC are shown in Figures \ref{fig:GenPass_Eden} and \ref{fig:GenPass_NEC}.
For each passenger volume, the integrated Eulerian energy density and NEC violations are used to compare how well each transport the same passenger volume at the same velocity. The results of this are shown in Table \ref{tab:PassResultsSum}. The change in passenger volume shape does see an improvement in the amount of energy required and a reduction in the NEC violation, with the pancake and teardrop shapes performing the best. However, these improvements are all under an order of magnitude, which means the selection of passenger volume is not the primary factor in either physicality nor massively reducing the needed energy. In Figure \ref{fig:GenPass_Scatter}, the impact of the different passenger volume shapes on the density and integrated Eulerian energy and NEC violations are shown.

\begin{table}[h]
    \centering
    \caption{Evaluation results for the different passenger volumes. The total Eulerian energy is a pointwise sum of the locally transformed stress-energy tensor $T^{00}$ component across each point in spacetime, making this value a representation of the sum total of each perceived energy, for each observer, at each point. This value is not the total integrated energy as viewed from a single observer across spacetime.}
    \begin{tabular}{|p{3cm}|c|c|c|c|c|c|}
        \multicolumn{1}{p{3cm}}{Parameter} & \multicolumn{1}{c}{Spherical}  & \multicolumn{1}{c}{Pancake}  & \multicolumn{1}{c}{Cigar} & \multicolumn{1}{c}{Cylindrical} & \multicolumn{1}{c}{Bicone} & \multicolumn{1}{c}{Teardrop} \\
        \hline
        \textit{Total Eulerian Energy [J]} & $-4.2\times10^{42}$ & $-1.9\times10^{42}$ & $-6.3\times10^{42}$  & $-3.1\times10^{42}$  & $-3.8\times10^{42}$  & $-2.1\times10^{42}$ \\
        \hline
        \textit{Max Eulerian Energy Density [J/m$^3$]} & $-2.6\times10^{38}$ & $-2.0\times10^{38}$  & $-2.6\times10^{38}$   & $-2.6\times10^{38}$  & $-2.0\times10^{38}$  & $-0.9\times10^{38}$ \\
        \hline
        \textit{Total NEC Violation} & $-3.0\times10^{44}$ & $-2.0\times10^{44}$  & $-4.0\times10^{44}$   & $-2.5\times10^{44}$  & $-3.0\times10^{44}$  & $-2.4\times10^{44}$ \\
         \hline
         \textit{Max NEC Violation Density} & $-1.0\times10^{40}$ & $-0.7\times10^{40}$  & $-1.2\times10^{40}$   & $-0.9\times10^{40}$  & $-3.0\times10^{40}$  & $-2.3\times10^{40}$ \\
         \hline
    \end{tabular}
        \label{tab:PassResultsSum}
\end{table}

\clearpage

\subsection{Van Den Broeck Metric (VDM)}
\begin{figure}[hbt!]
\centering
\includegraphics[width=\textwidth]{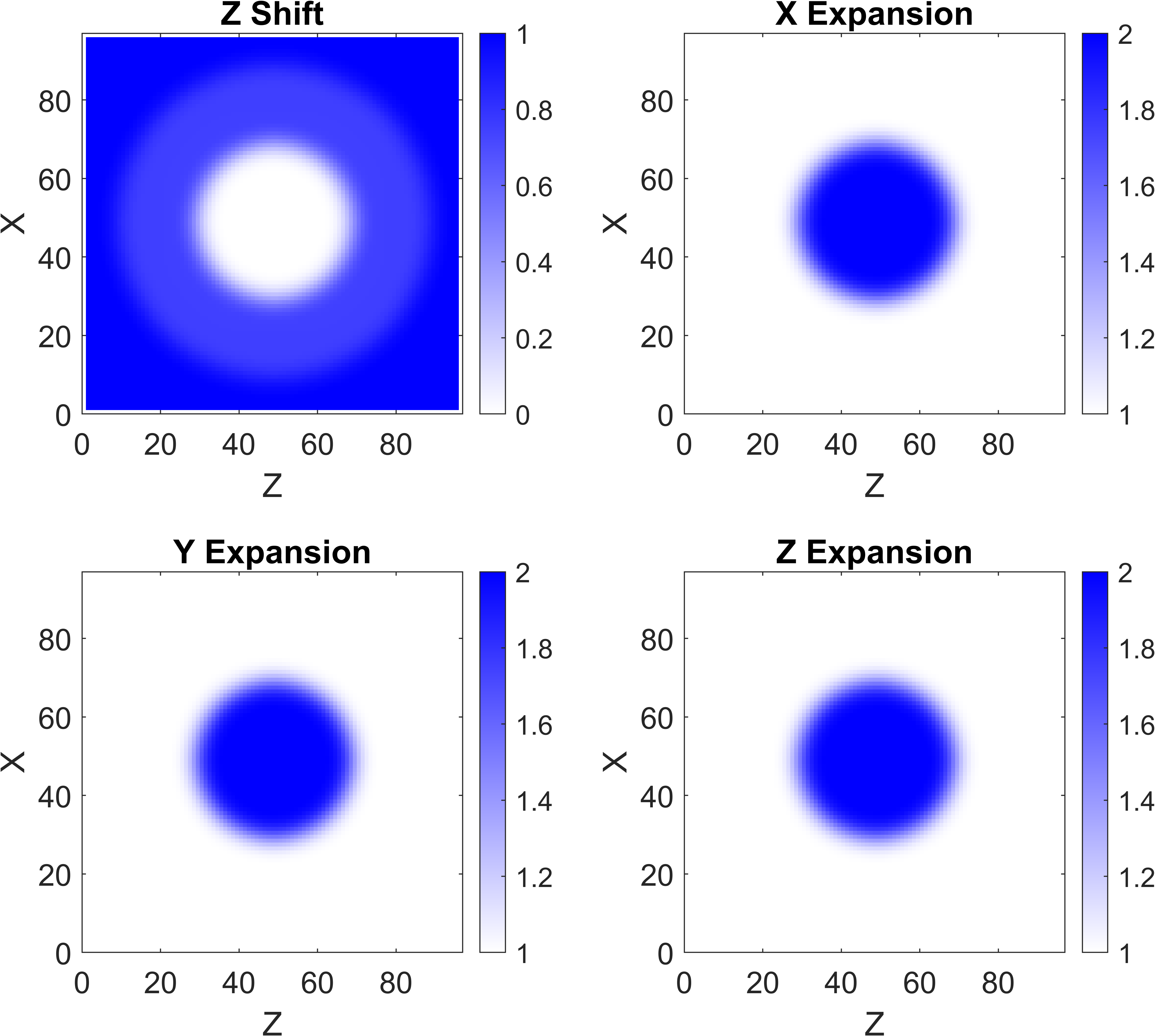}
\caption{Metric functions for the VDM. The VDM uses both the shift vector along Z and passenger volume expansion terms along X, Y, and Z. Cross-section is plotted for y = 0.}\label{fig:VDM_metric}
\end{figure}

\begin{figure}[hbt!]
\centering
\includegraphics[width=\textwidth]{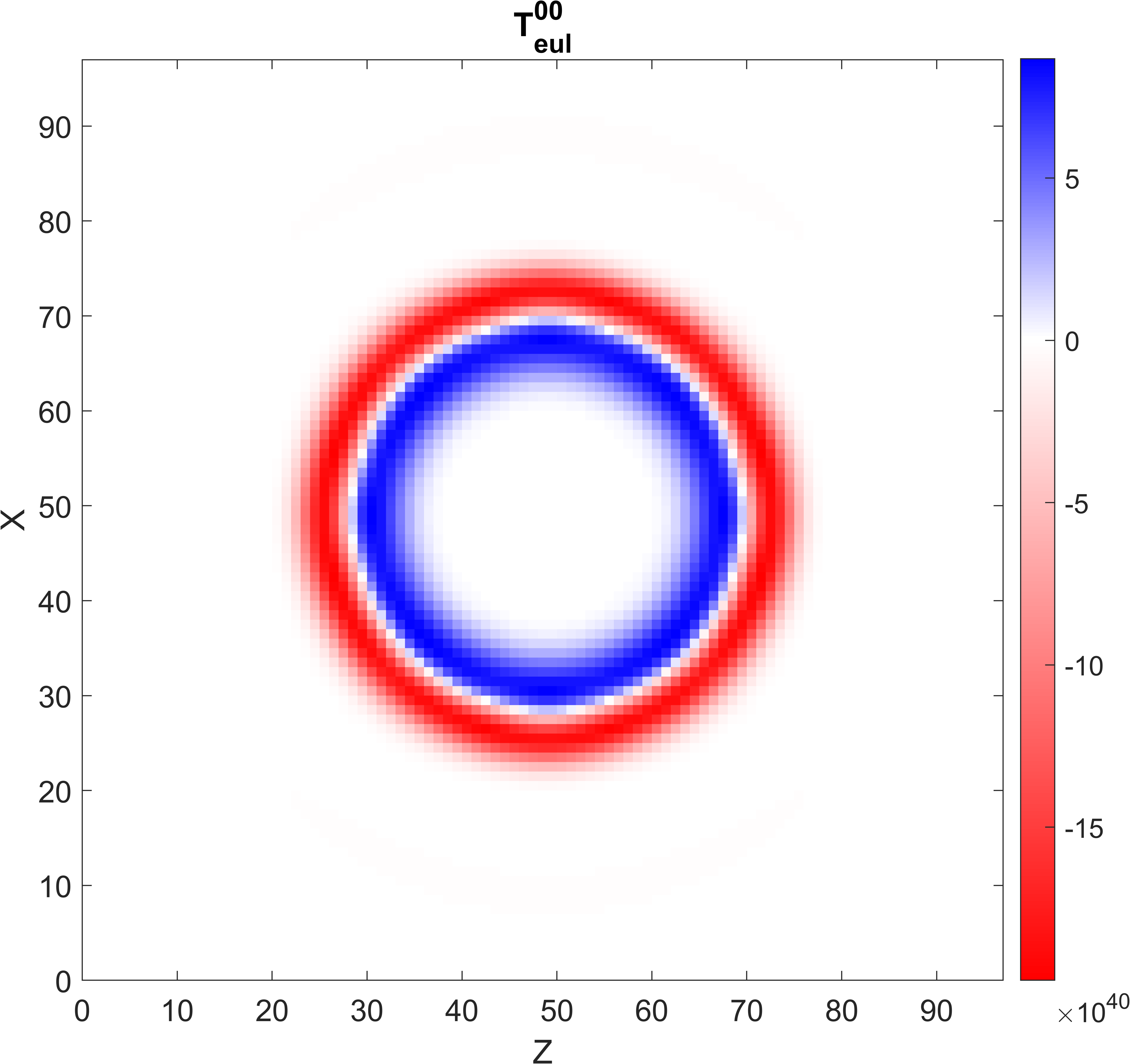}
\caption{Eulerian energy density for the VDM. Direction of motion is along +Z. Cross-section is plotted for y = 0.}\label{fig:VDM_Eden}
\end{figure}

\begin{figure}[hbt!]
\centering
\includegraphics[width=\textwidth]{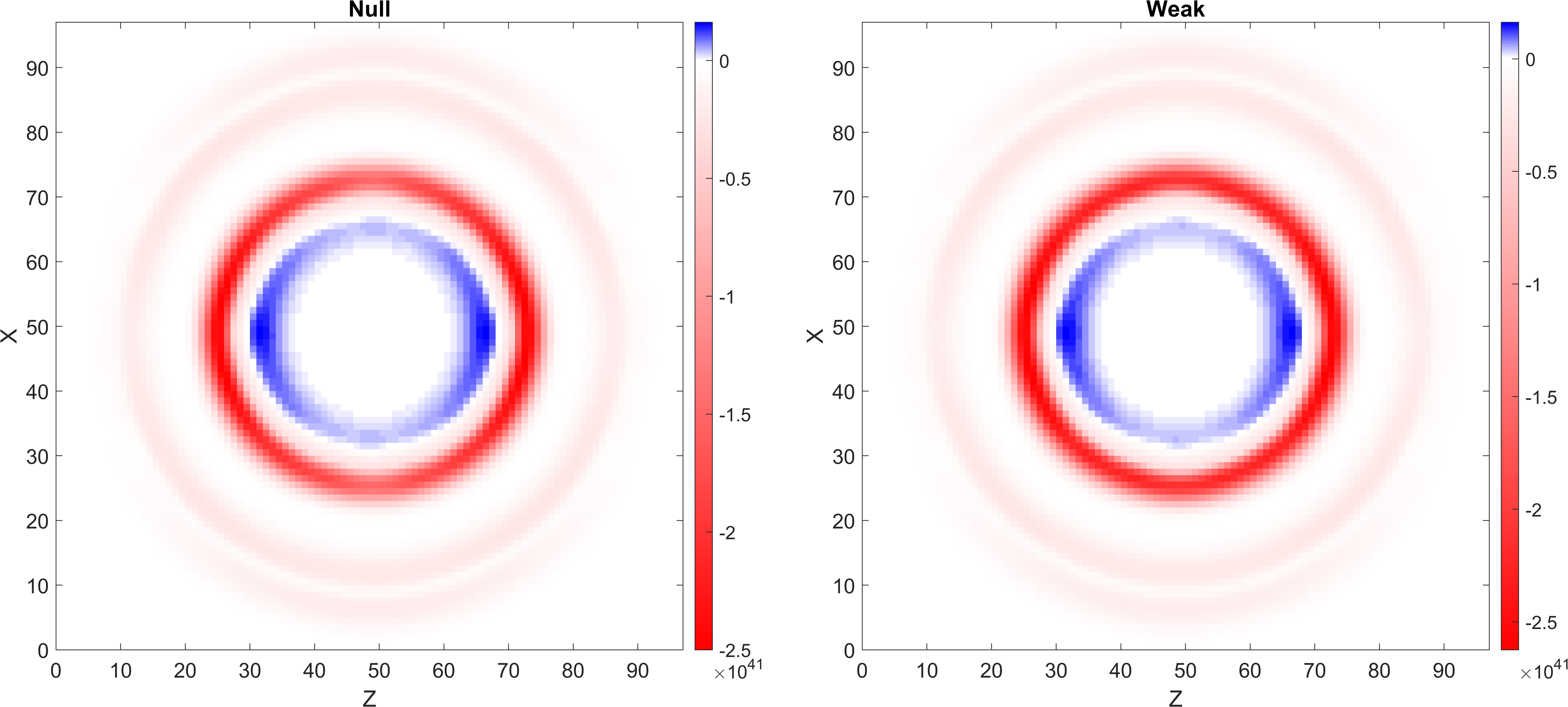}
\caption{NEC and WEC for the VDM. Direction of motion is along +Z. Plotted are the most violating values across all observers. When only positive energy density is seen across all null or timelike observers, then the minimum positive value is shown. Cross-section is plotted for y = 0.}\label{fig:VDM_Econd}
\end{figure}

\begin{figure}[hbt!]
\centering
\includegraphics[width=\textwidth]{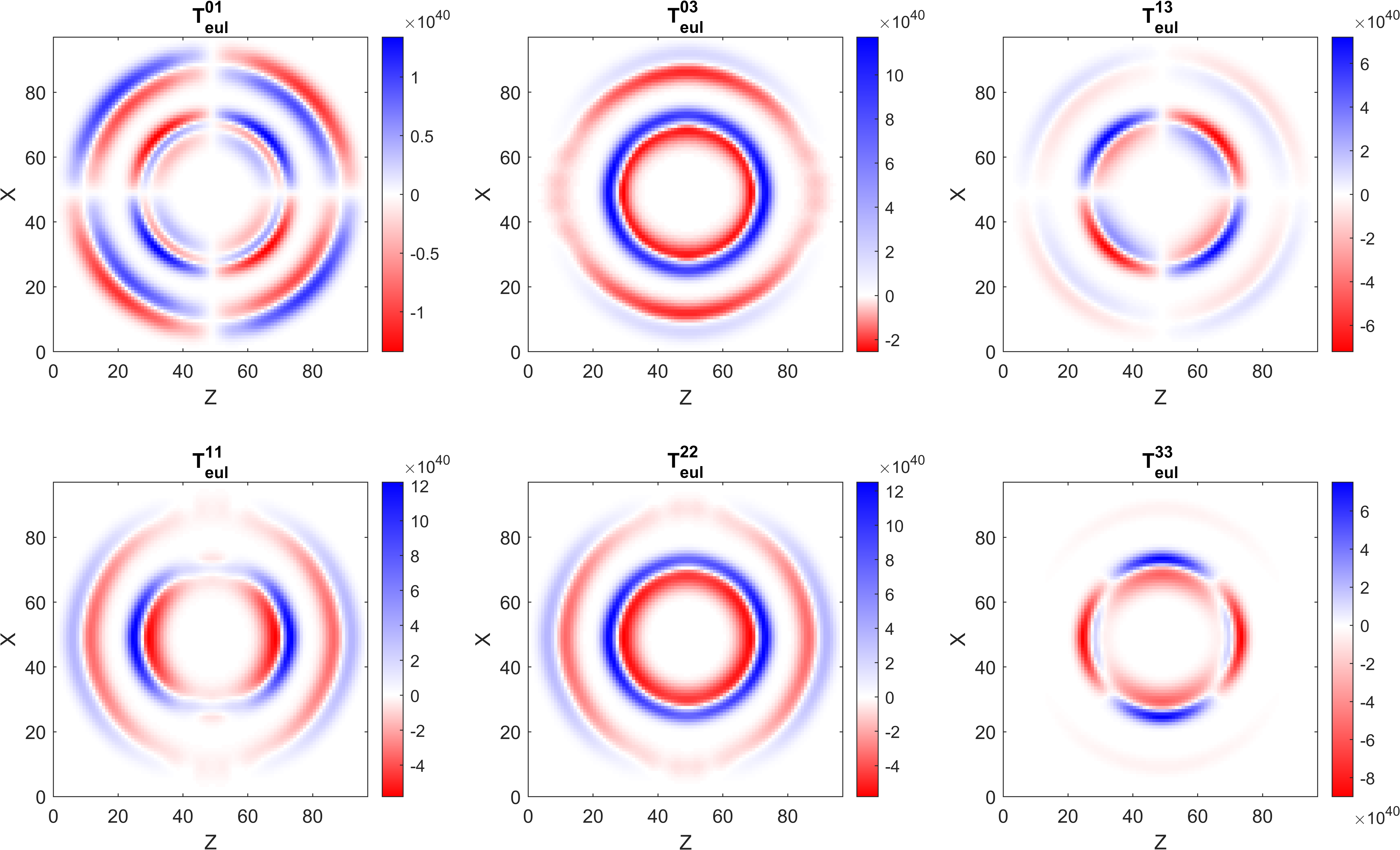}
\caption{Eulerian stress-energy tensor for the VDM. Direction of motion is along +Z. Cross-section is plotted for y = 0.}\label{fig:VDM_Etensor}
\end{figure}

Following Alcubierre's publication, Van Den Broeck built a modification to the AM in 1999\cite{1999CQGra..16.3973V} by using concentric regions which varied the shift and expansion components in an effort to reduce the negative energy required via the expansion of the passenger volume. The metric components cross-section is shown in Figure \ref{fig:VDM_metric} demonstrating these regions in the Z shift vector value. 
The Eulerian energy density has a similar shape to that of Alcubierre but has a positive region on the inside through the transition into the passenger volume with its expansion. This is shown in Figure \ref{fig:VDM_Eden}. Not surprisingly, this metric has WEC violations similar to Alcubierre, shown in Figure \ref{fig:VDM_Econd}. The existence of negative energy in the Eulerian frame is sufficient to know that a violation exists but the reason for different rings of violation is better understood when considering the stress-energy tensor, shown in Figure \ref{fig:VDM_Etensor}. The interaction of the pressure and momentum density terms when the energy density is small causes the main violation regions.

\begin{figure}[hbt!]
\centering
\includegraphics[width=\textwidth]{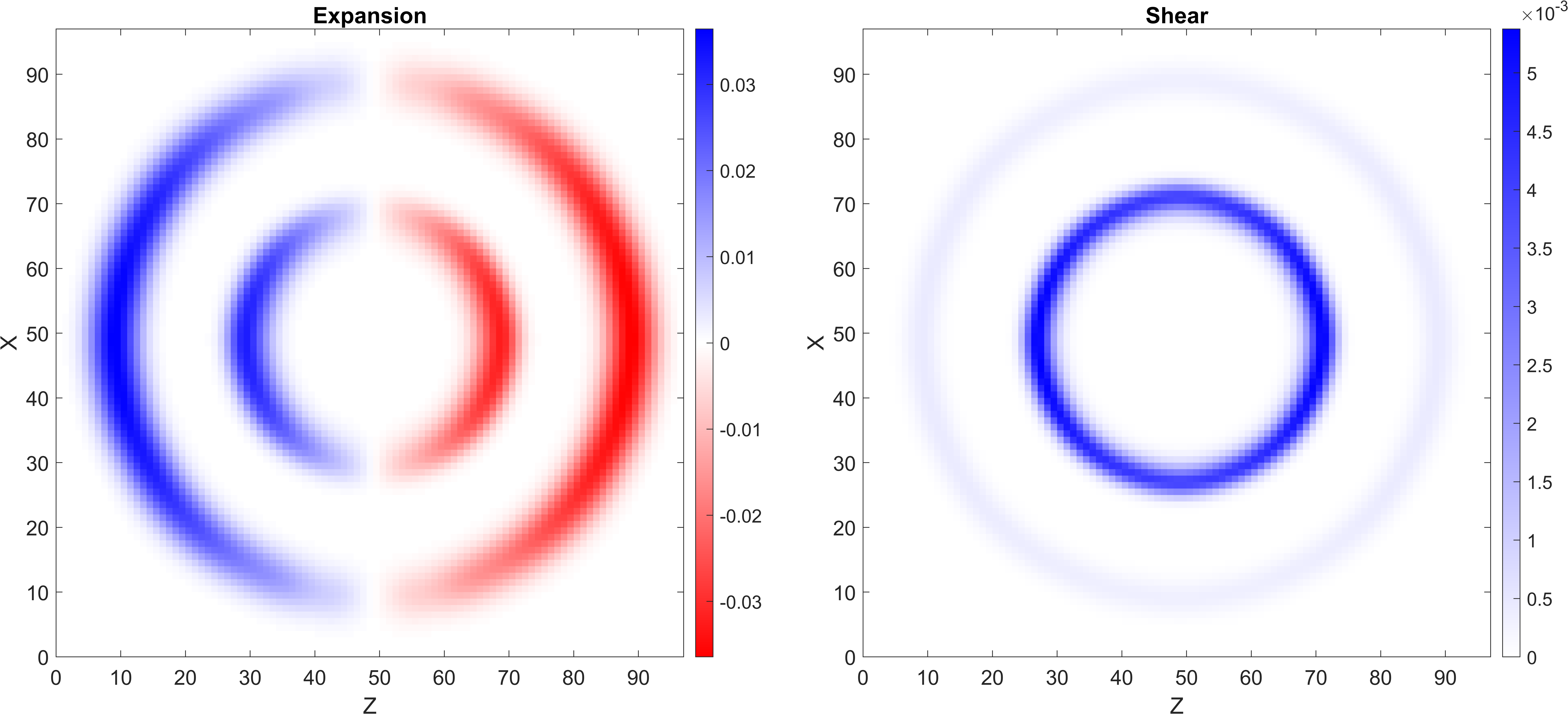}
\caption{The expansion and shear scalars for the VDM. Direction of motion is along +Z. Cross-section is plotted for y = 0.}\label{fig:VDM_scalars}
\end{figure}

Finally, the expansion and shear scalars are shown in Figure \ref{fig:VDM_scalars}. Each of the transition regions between the passenger volume and bubble has expansion and shear present, in a form similar to the AM. 

Ultimately, the VDM does not provide an approach that resolves WEC violation, but it does provide a novel method to make a warp drive more efficient, as the expansion terms allow more passenger volume per required energy.
\clearpage

\subsection{Bobrick-Martire Modified Time Metric (MTM)}
\begin{figure}[hbt!]
\centering
\includegraphics[width=\textwidth]{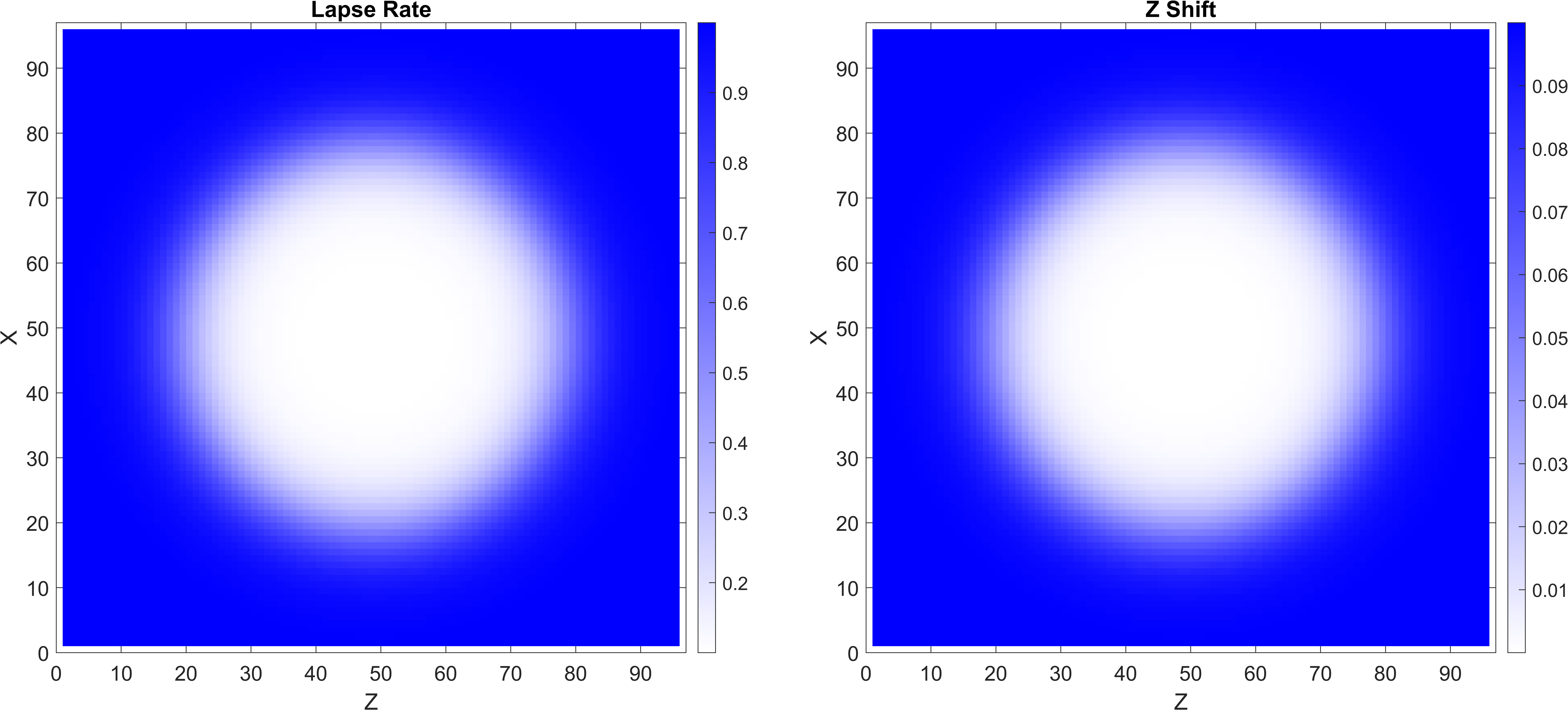}
\caption{Metric functions for the MTM. Only the lapse and shift vector along Z have non-Minkowski values. Direction of motion is along +Z. Cross-section is plotted for y = 0.}\label{fig:BM_functions}
\end{figure}

\begin{figure}[hbt!]
\centering
\includegraphics[width=\textwidth]{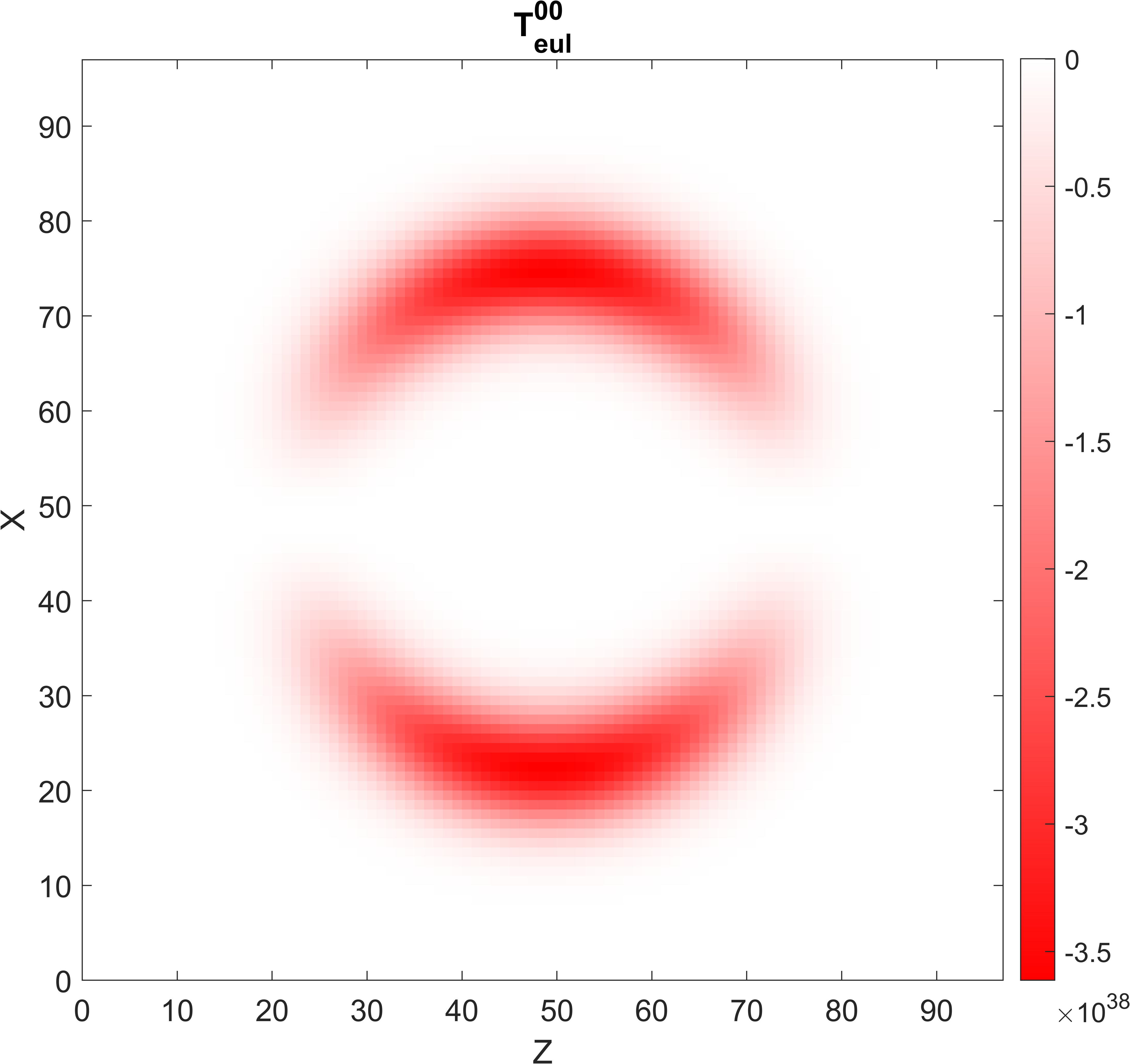}
\caption{Eulerian energy density for the MTM. Cross-section is plotted for y = 0.}\label{fig:BM_Eden}
\end{figure}

\begin{figure}[hbt!]
\centering
\includegraphics[width=\textwidth]{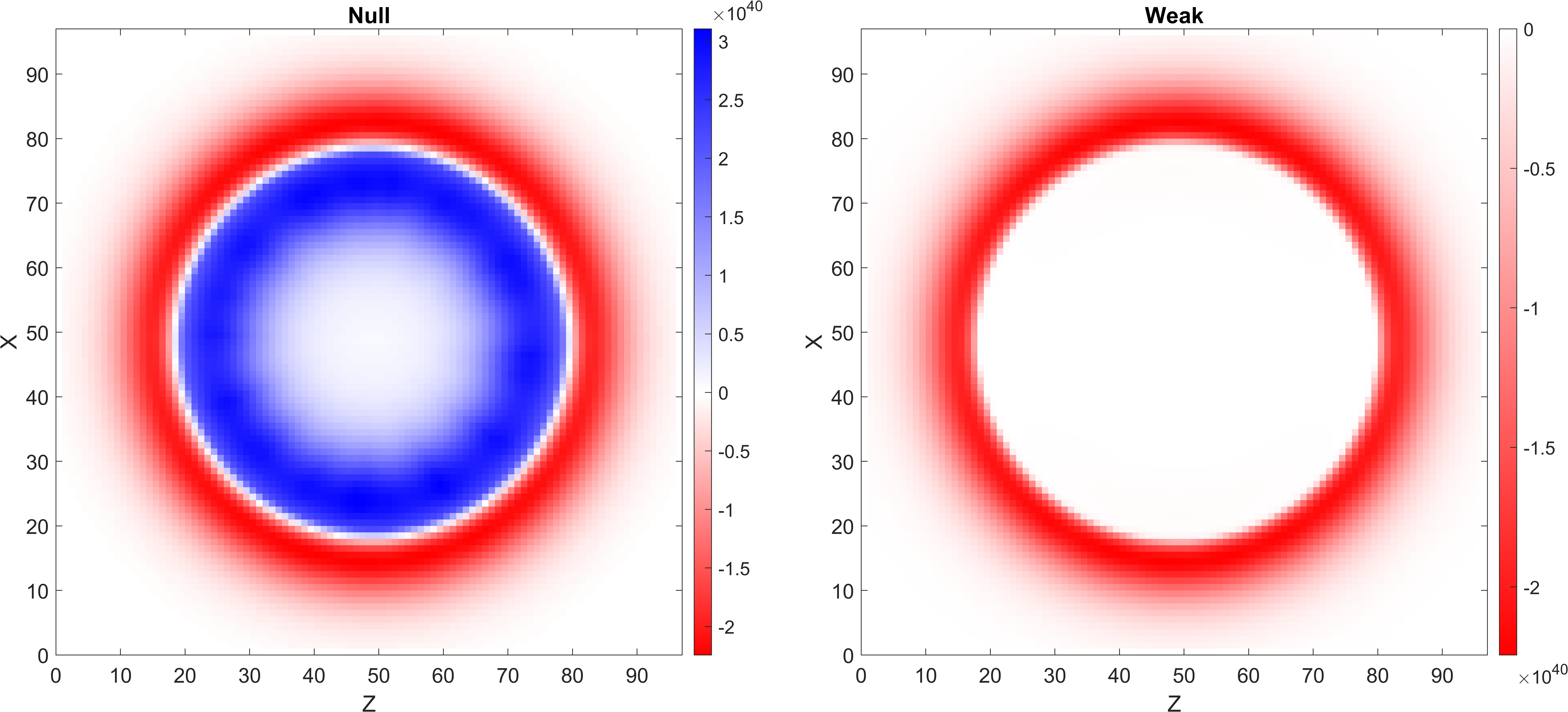}
\caption{NEC and WEC for the MTM. Direction of motion is along +Z. Plotted are the minimum values across all observers. When only positive energy density is seen across all null or timelike observers, the minimum will be positive. Cross-section is plotted for y = 0.}\label{fig:BM_Econd}
\end{figure}

\begin{figure}[hbt!]
\centering
\includegraphics[width=\textwidth]{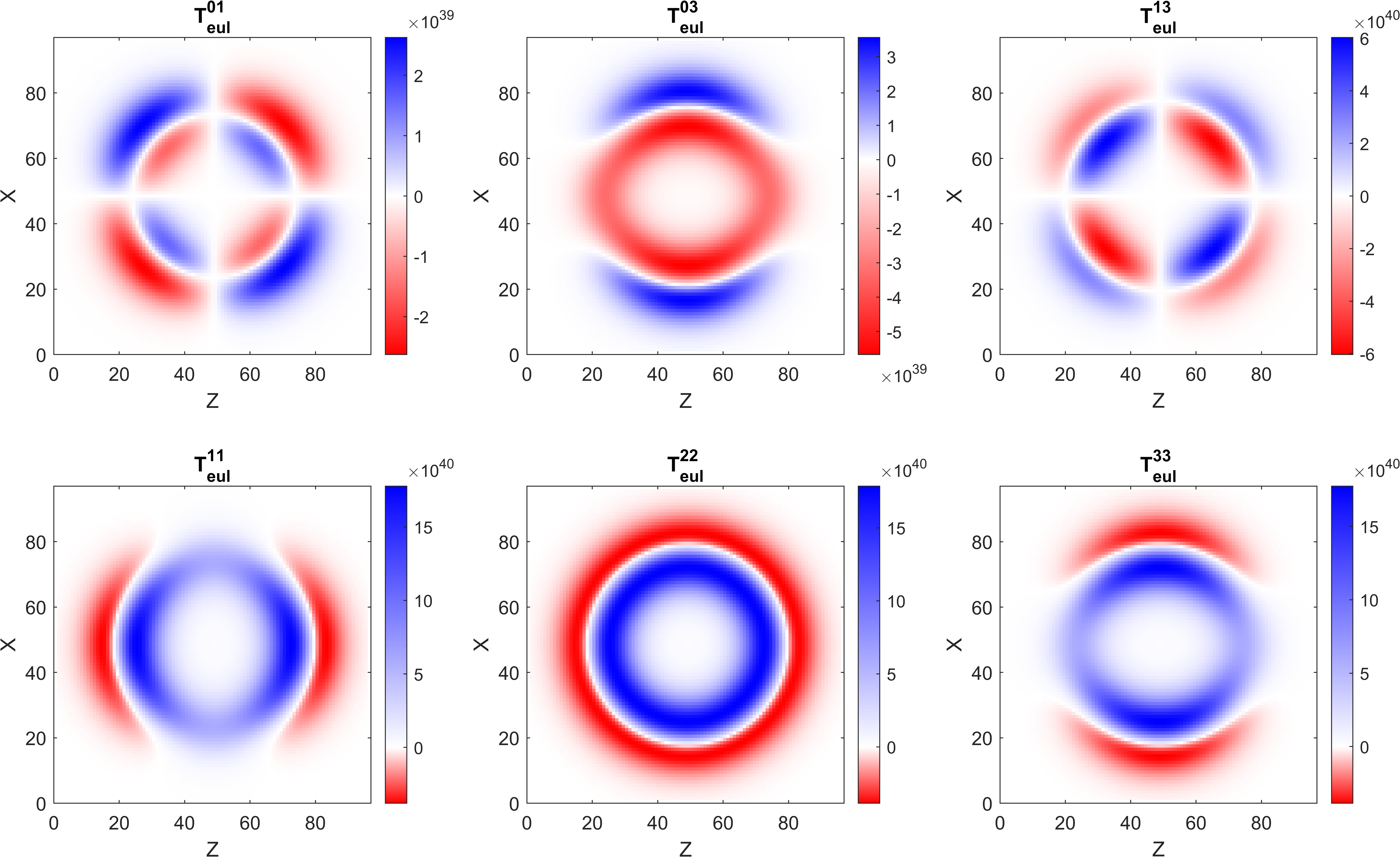}
\caption{Eulerian stress-energy tensor components for the MTM. The $T^{00}$ term is shown in Figure \ref{fig:BM_Eden}. Cross-section is plotted for y = 0.}\label{fig:BM_Etensor}
\end{figure}

\begin{figure}[hbt!]
\centering
\includegraphics[width=\textwidth]{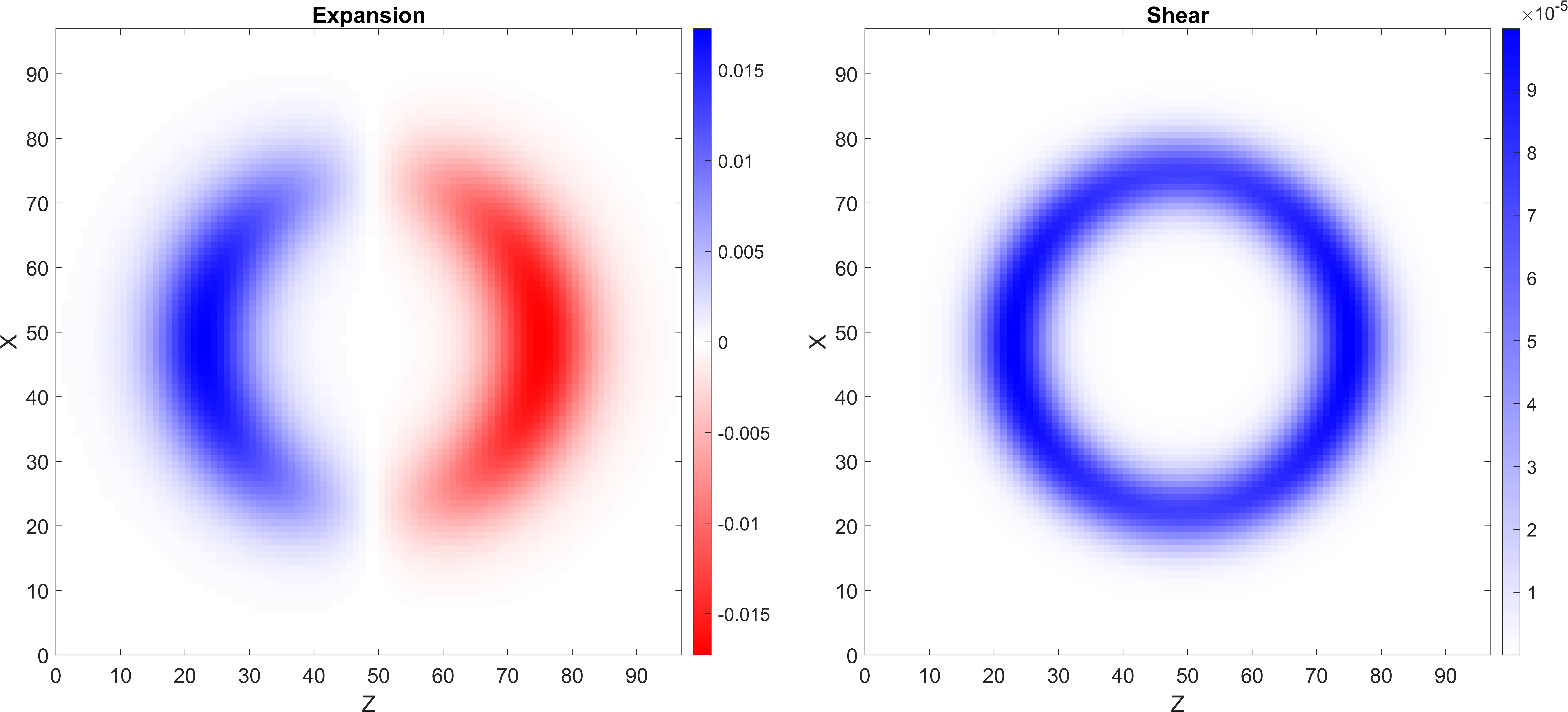}
\caption{Expansion and shear scalars for the MTM. Direction of motion is along +Z. Cross-section is plotted for y = 0.}\label{fig:BM_scalars}
\end{figure}

In Introducing Physical Warp Drives \cite{2021CQGra..38j5009B}, Bobrick and Martire consider several modifications to the AM, one of these being the addition of a changing lapse rate (see section 4.5 in \cite{2021CQGra..38j5009B}). The metric functions for this kind of drive are shown in Figure \ref{fig:BM_functions}. The impact of a changing lapse is a reduction in the passage of time for passengers inside the drive as compared to an outside observer.
The Eulerian energy density resulting from an addition of changing passenger lapse is remarkably similar to Alcubierre, with the energy density shown in Figure \ref{fig:BM_Eden}. However, when considering the energy conditions, a much different result is found, shown in Figure \ref{fig:BM_Econd}. For both NEC and WEC there is no longer any violation inside the inner region, with the NEC only seeing positive values within that region. Looking at the stress-energy tensor in Figure \ref{fig:BM_Etensor}, positive pressures are seen on the inner region of the bubble, as opposed to the negative values seen in the AM.
Finally, the expansion and shear scalars are shown in Figure \ref{fig:BM_scalars}. These share the same shape and values as seen in the AM.

While the addition of a changing lapse to the metric is not a fix for all violations of the WEC, it does offer an improvement over the AM by removing violations in the inner region of the bubble.

\clearpage

\subsection{Lentz Metric (LM)}
\begin{figure}[hbt!]
\centering
\includegraphics[width=\textwidth]{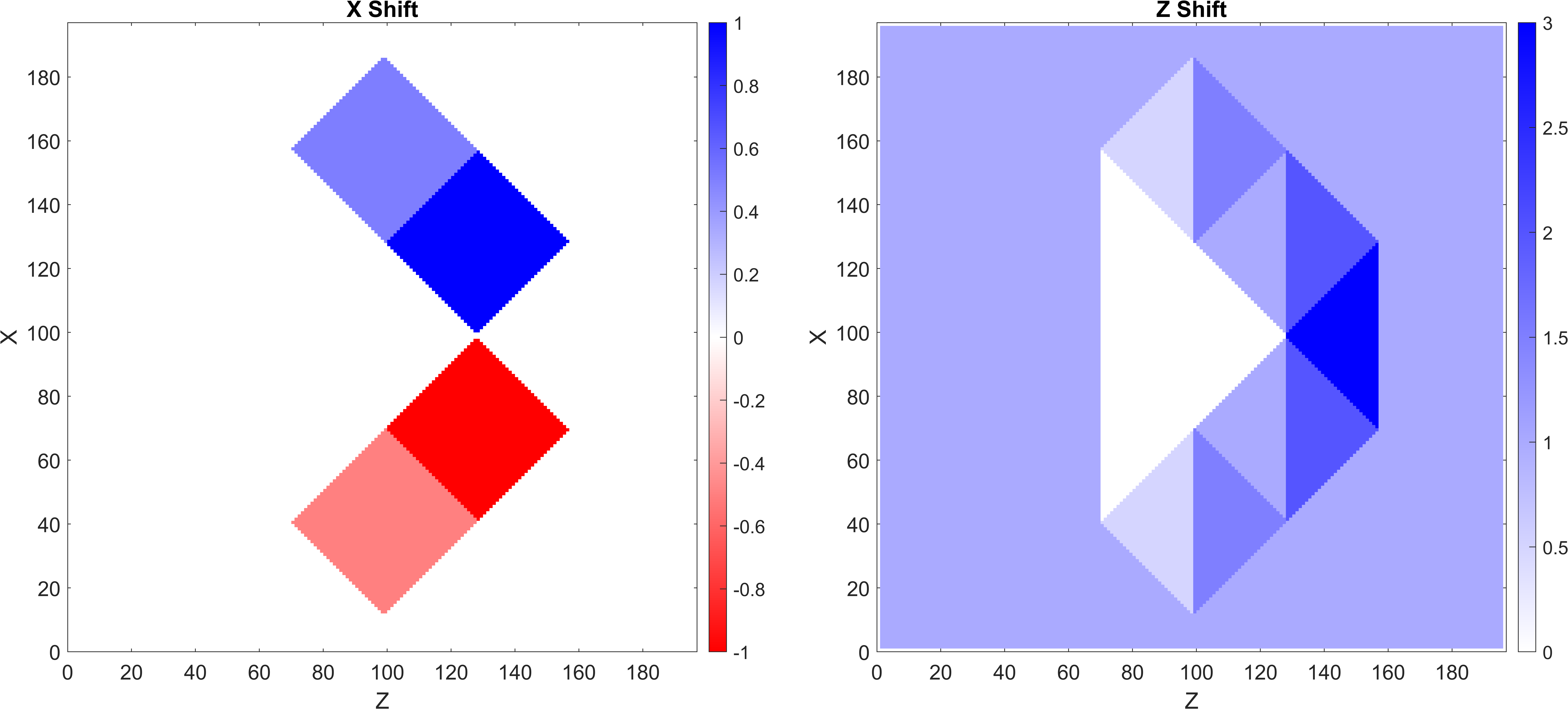}
\caption{Metric functions for the LM. Only the shift vectors along X and Z have values different from Minkowski. These functions are shown in the comoving frame. Direction of motion is along +Z. Cross-section is plotted for y = 0.}\label{fig:LM_functions}
\end{figure}

\begin{figure}[hbt!]
\centering
\includegraphics[width=\textwidth]{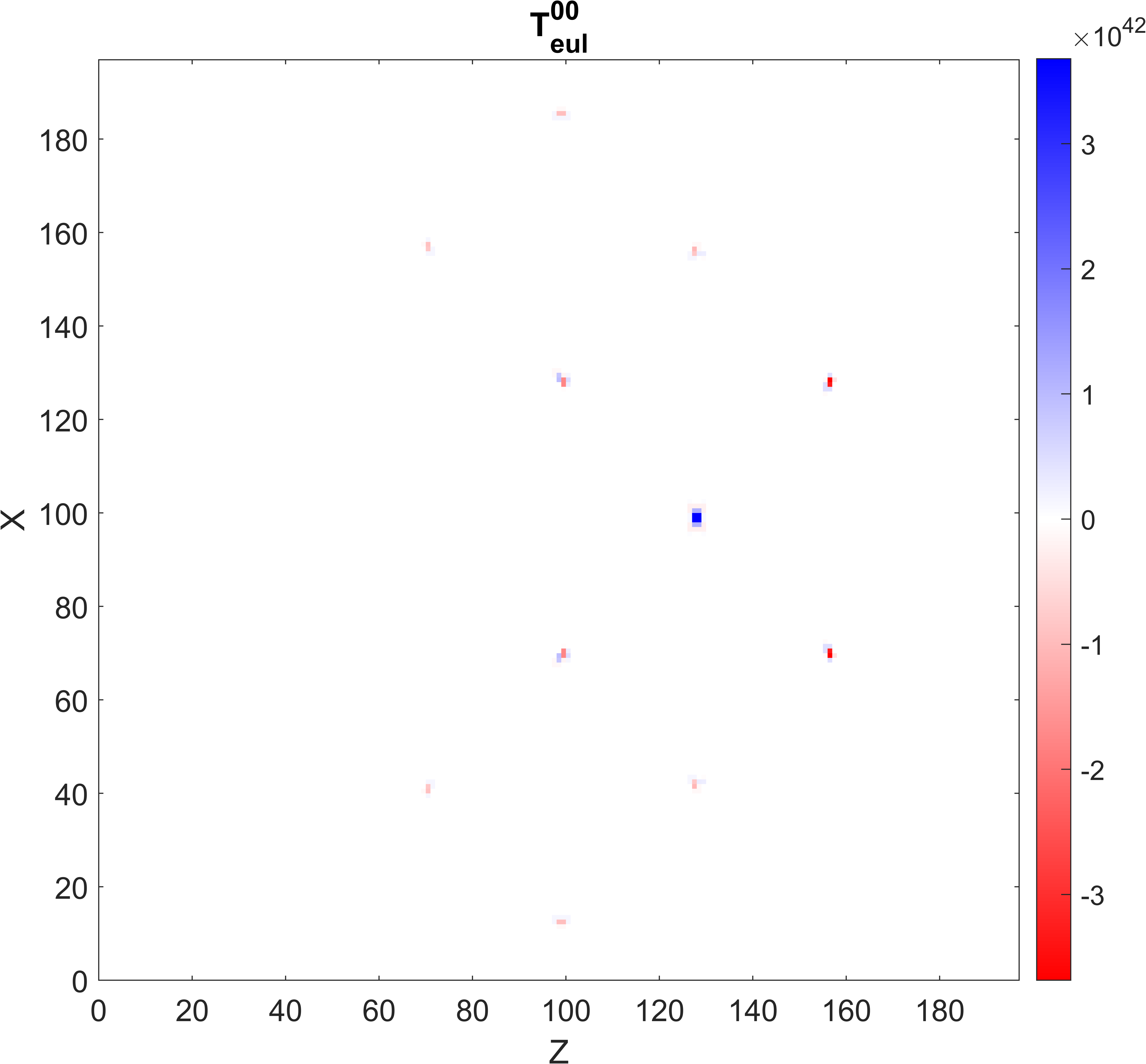}
\caption{Eulerian energy density for the LM. Each point of energy density aligns with the corners of the rhomboid shape of the shift vector. Direction of motion is along +Z. Cross-section is plotted for y = 0.}\label{fig:LM_Eden}
\end{figure}

\begin{figure}[hbt!]
\centering
\includegraphics[width=\textwidth]{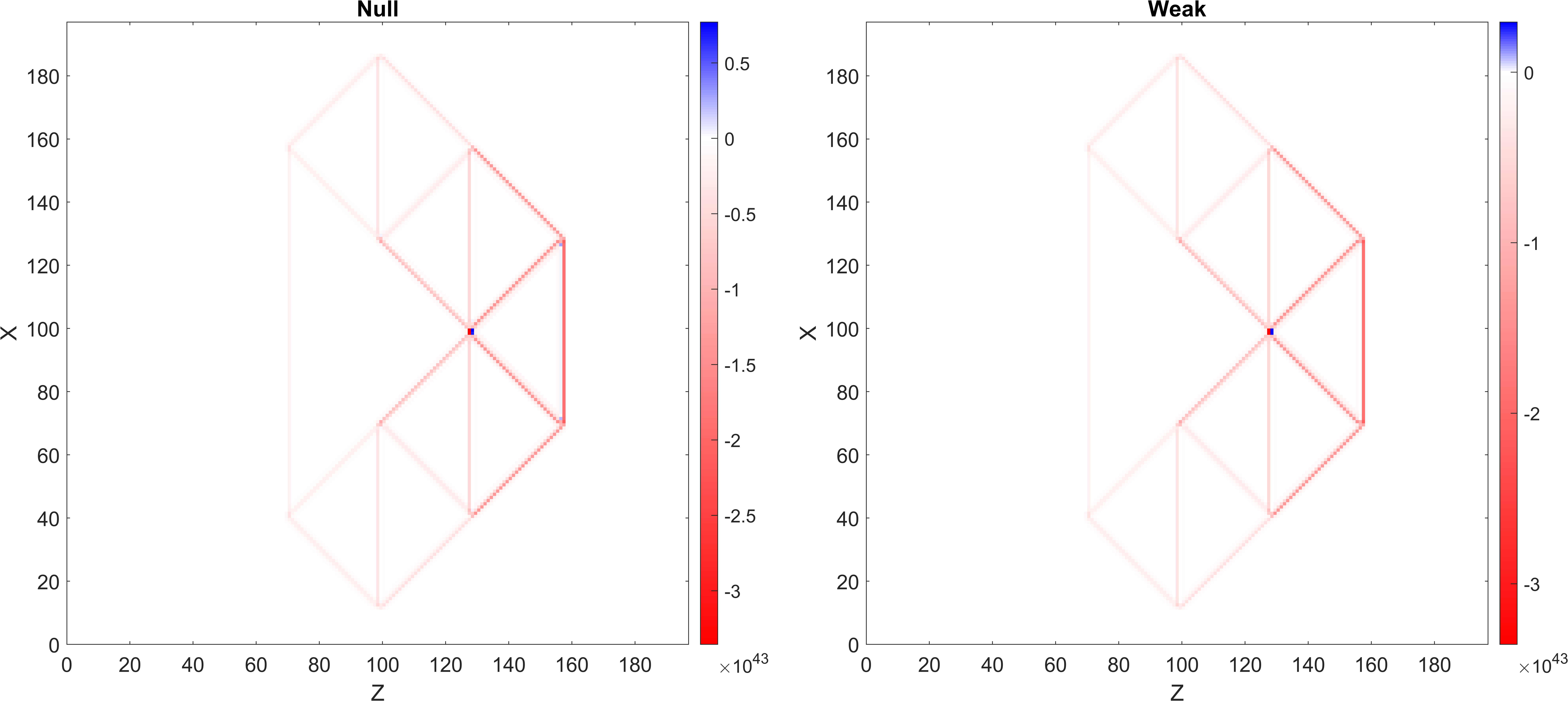}
\caption{NEC and WEC for the LM. Direction of motion is along +Z. Plotted are the minimum values across all observers. When only positive energy density is seen across all null or timelike observers, the minimum positive value is shown. Cross-section is plotted for y = 0.}\label{fig:LM_Econd}
\end{figure}

\begin{figure}[hbt!]
\centering
\includegraphics[width=\textwidth]{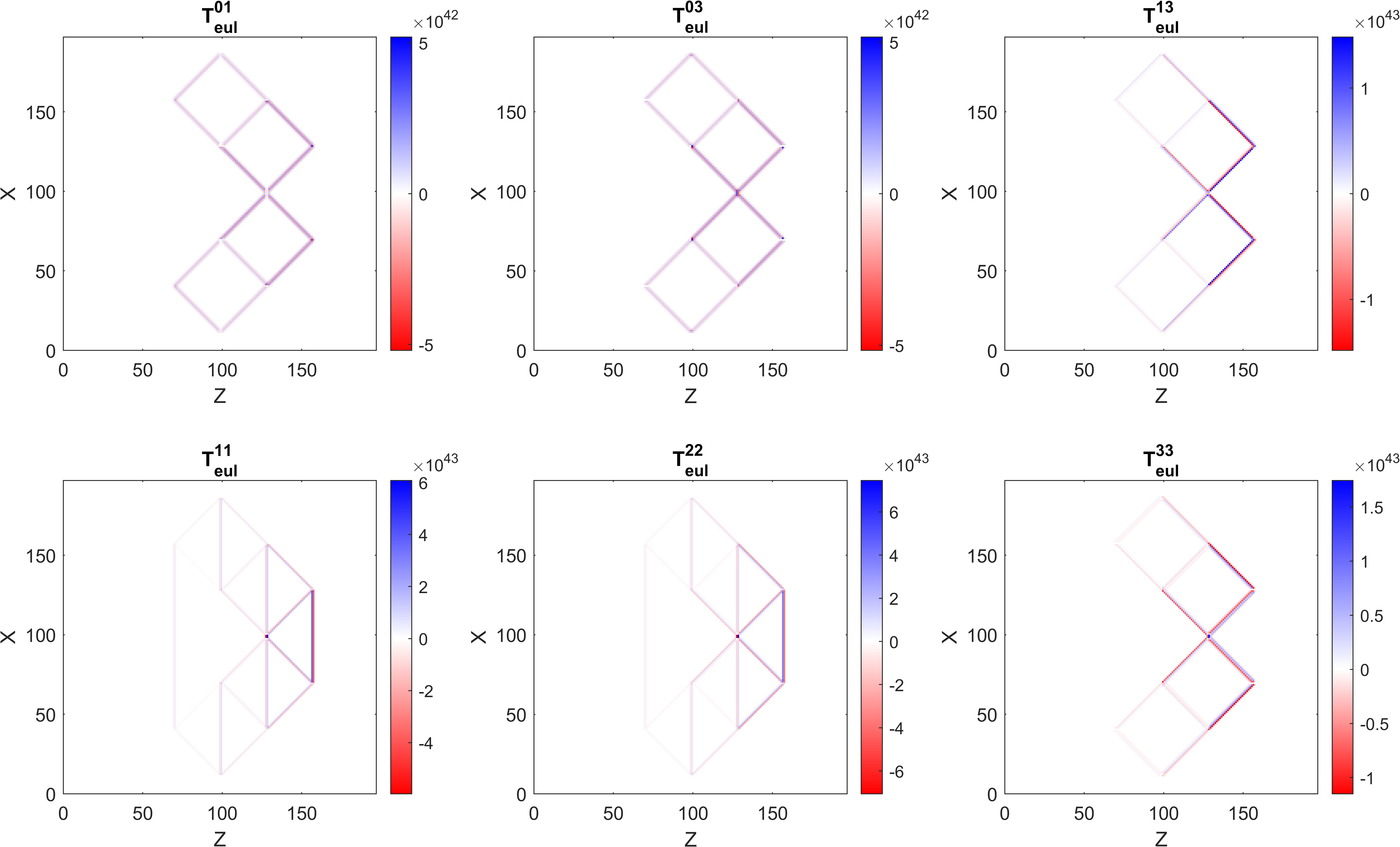}
\caption{The non-zero stress-energy tensor components, in the Eulerian frame, for the LM. The $T^{00}$ component is shown in Figure \ref{fig:LM_Eden}. Direction of motion is along +Z. Cross-section is plotted for y = 0.}\label{fig:LM_tensor}.
\end{figure}

The Lentz metric (LM) \cite{2022arXiv220100652L} is a radical departure from "top-hat" like spherically symmetric solutions proposed by Alcubierre and expanded on by Van Den Broeck, Bobrick, and Martire. The Lentz solution proposes to avoid WEC violations entirely, even in superluminal regimes. This is accomplished by recognizing that one issue with the prior warp metrics is the use of a single shift vector component, which provides a source of the negative energy.  Lentz proposes that additional shift components can be used to construct a "potential" in the form of a linear wave equation:
\begin{equation}
    \partial_x^2 \phi+\partial_y^2 \phi-\frac{2}{v_h^2} \partial_z^2 \phi=\rho
\end{equation}
which is related to the shift vector as
\begin{equation}
    N_i = \partial_i \phi
\end{equation}
and related to the Eulerian energy density by:
\begin{equation}\label{eq:LentzE}
    E=\frac{1}{16 \pi}\left(2 \partial_z^2 \phi\left(\rho+\frac{2}{v_h^2} \partial_z^2 \phi\right)-4\left(\partial_z \partial_x \phi\right)^2\right)
\end{equation}
A proposed solution to Equation \ref{eq:LentzE} is a rhomboid-like shift vector along any two dimensions (in Cartesian coordinates). The APL team has built a version of this metric and evaluated it using the Warp Factory. Our form of the LM shift vector is shown in Figure \ref{fig:LM_functions}.
The APL evaluation of the energy density, shown in Figure \ref{fig:LM_Eden}, shows a similar result from Lentz's original work \cite{2022arXiv220100652L}, with no major points of the energy density outside of the corners. This result differs from Lentz in that negative energy does exist for a few points, in addition to positive energy. However, this is likely due to the smoothing used in the shift vectors around the edges\footnote{The smoothing applied by Lentz is unclear from his paper, but it is likely the case that the negative energy at the corners can be removed by more clever management of the gradients.}. While the Eulerian energy density is mostly zero, the evaluations of the NEC and WEC show that violations occur throughout the structure, shown in Figure \ref{fig:LM_Econd}. The reason for this violation can be seen by taking a closer look at the stress-energy tensor, shown in Figure \ref{fig:LM_tensor}.

\begin{figure}[hbt!]
\centering
\includegraphics[width=\textwidth]{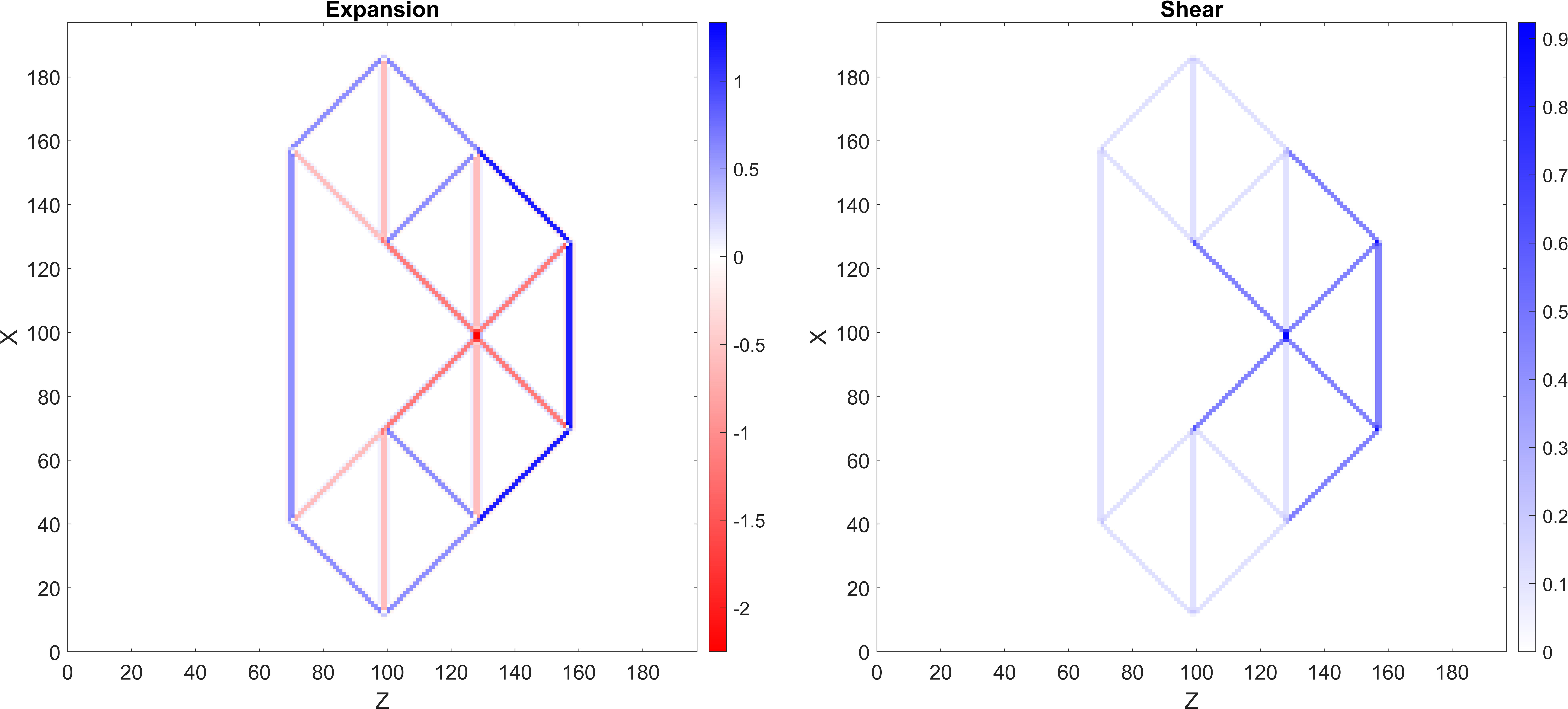}
\caption{Expansion and shear scalars for the LM. Direction of motion is along +Z. Cross-section is plotted for y = 0.}\label{fig:LM_scalars}
\end{figure}

The culprit for WEC violation is the existence of Eulerian pressure terms at all shift vector boundaries with an absence of energy density. When transformed into other timelike observer frames, this appears as negative energy density to that observer. Finally, the scalars for the LM are shown in Figure \ref{fig:LM_scalars}. 

Lentz's proposed soliton solution does provide a unique metric form that differs greatly from the "top-hat" form of Alcubierre and removes the major violations of Eulerian energy density. However, when running a full analysis across different timelike observers, this solution still results in WEC violations. 
\clearpage

\subsection{Summary}
The prior results of evaluating common and newly proposed metrics using Warp Factory are summarized in this section in terms of the statements of physicality (see Section \ref{sec:physicality}). The results for each metric are shown in Table \ref{tab:PhySum}.
\begin{table}[h]
    \centering
    \caption{A summary of the physicality conditions of the warp drive metrics discussed and evaluated in this paper. For the LM, the smoothing applied in this work does see regions of negative Eulerian energy, but this is due to not being able to replicate the smoothing applied in Lentz's paper. For the VDM, the reasonable mass condition could possibly be met, but requires very small regions with very large expansions which have not yet been analyzed by our code.}
    \begin{tabular}{|b{5cm}|c|c|c|c|}
        
        \multicolumn{1}{b{4.6cm}}{Conditions} & \multicolumn{1}{>{\centering\arraybackslash} p{2.3cm}}{Alcubierre \ \ \ \  Metric} & \multicolumn{1}{>{\centering\arraybackslash}  p{2.3cm}}{Van-Den Broeck Metric} & \multicolumn{1}{>{\centering\arraybackslash} 
 p{2.3cm}}{Bobrick-Martire Modified Time}  & \multicolumn{1}{>{\centering\arraybackslash} p{2.3cm}}{Lentz \ \ \ \ \ \ \  Metric}  \\
        \hline
        \textit{Non-negative Eulerian  energy} & \cellcolor{red} \ding{53}  & \cellcolor{red} \ding{53} & \cellcolor{red} \ding{53}  & \cellcolor{yellow} ?  \\
        \hline
        \textit{Satisfies NEC everywhere} & \cellcolor{red} \ding{53} & \cellcolor{red} \ding{53} & \cellcolor{red} \ding{53} &  \cellcolor{red} \ding{53}  \\
         \hline
        \textit{Satisfies WEC everywhere} & \cellcolor{red} \ding{53} & \cellcolor{red} \ding{53} & \cellcolor{red} \ding{53} &\cellcolor{red} \ding{53}  \\
        \hline
        \textit{Uses reasonable mass} & \cellcolor{red} \ding{53} & \cellcolor{yellow} ? & \cellcolor{red} \ding{53} & \cellcolor{red} \ding{53} \\
        \hline
    \end{tabular}
        \label{tab:PhySum}
\end{table}
Using Warp Factory, elements of the stress-energy tensor can be fully explored to understand better how NEC and WEC violations occur. Across all of the metrics discussed in this paper, the common source of energy condition violation appears to stem from either the existence of negative energy density (as evaluated by a typical Eulerian observer) or the existence of pressure and momentum density terms exceeding the energy density at certain points. Once transformed into other timelike observers, these other terms become a negative energy seen by these other observers. 

This leads to an important insight for warp analysis. To properly check for physicality, a full sampling of the null and timelike observers must be done to establish if any violation occurs. Determining only an Eulerian observer's energy density as non-negative is not a sufficient condition when discussing physicality.

\clearpage

\section{Optimization Results}\label{sec:Optimization}
The optimization method, discussed in Section \ref{sec:OptModule}, can take a starting metric and perturb it to generate an entirely new metric. This is a computationally intensive process but does offer some insight into what changes to a metric tend to improve physicality. To demonstrate the optimization capability a starting AM is used with a passenger volume radius of 20 m and a velocity in the z direction of 0.5c\footnote{The bubble parameter $\sigma = 0.1$}. A genetic algorithm is run, with parameters described in Table \ref{tab:OptStartData}. These optimizations are performed on an axisymmetric 2D coordinate system and translated following the optimization into 3D for visualization.

\begin{figure}[hbt!]
\centering
\includegraphics[width=\textwidth]{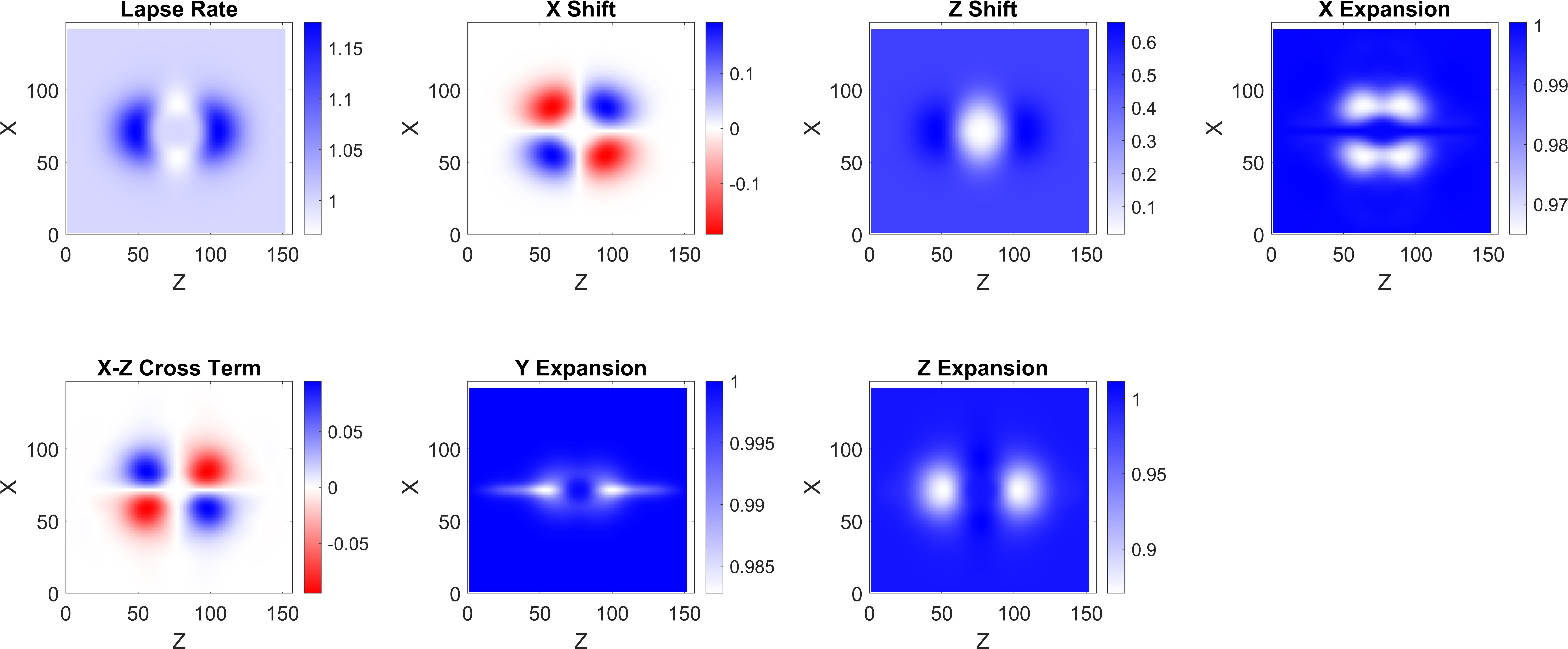}
\caption{Metric functions after optimization. Direction of motion is along +Z. Cross-section is plotted for y = 0.}\label{fig:Opt_Functions}
\end{figure}

\begin{figure}[hbt!]
\centering
\includegraphics[width=\textwidth]{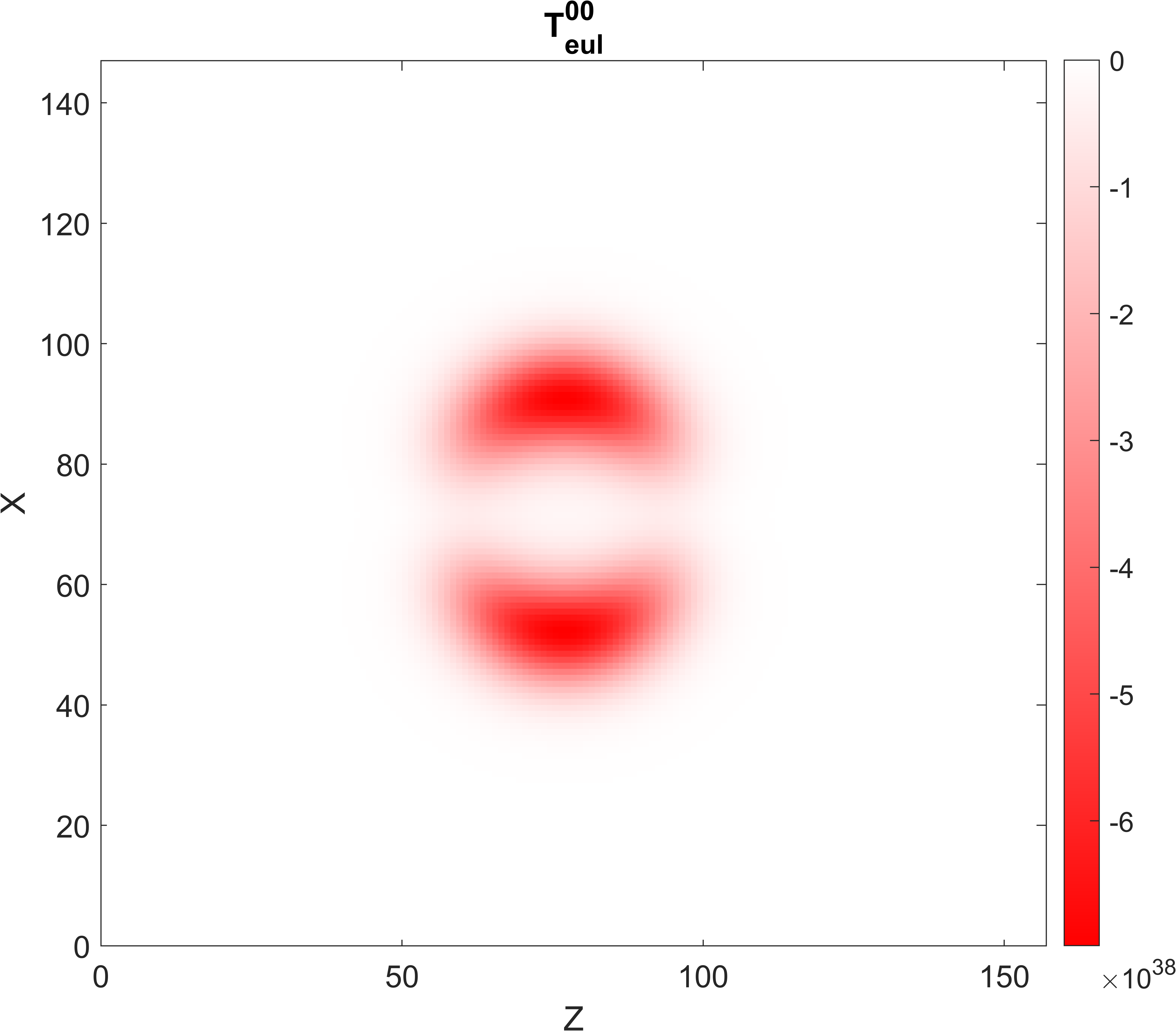}
\caption{Eulerian energy density of the starting metric. Direction of motion is along +Z. Cross-section is plotted for y = 0.}\label{fig:Opt_EdenStart}
\end{figure}

\begin{figure}[hbt!]
\centering
\includegraphics[width=\textwidth]{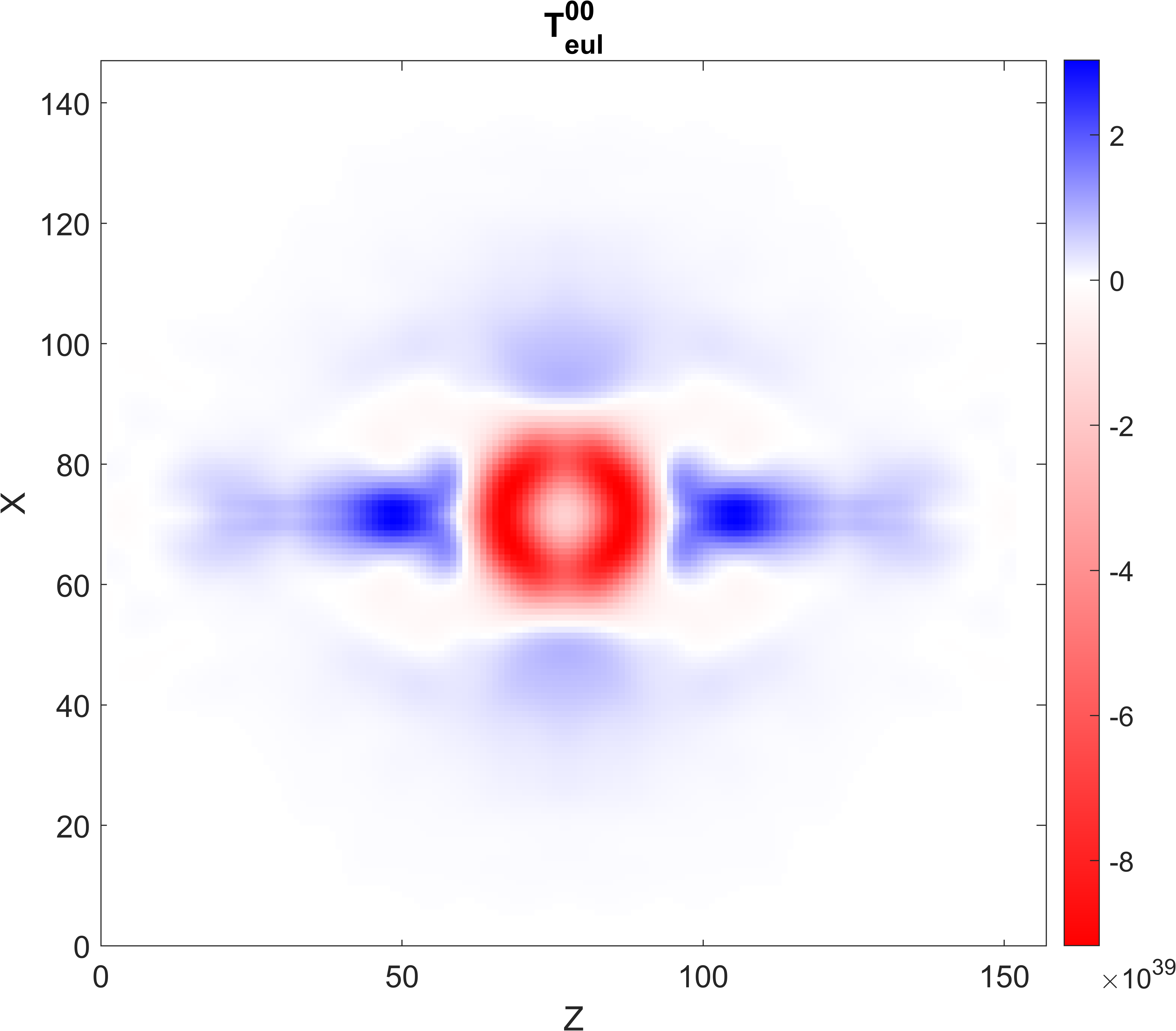}
\caption{Eulerian energy density of the optimized metric. Direction of motion is along +Z. Cross-section is plotted for y = 0.}\label{fig:Opt_EdenEnd}
\end{figure}

\begin{figure}[hbt!]
\centering
\includegraphics[width=\textwidth]{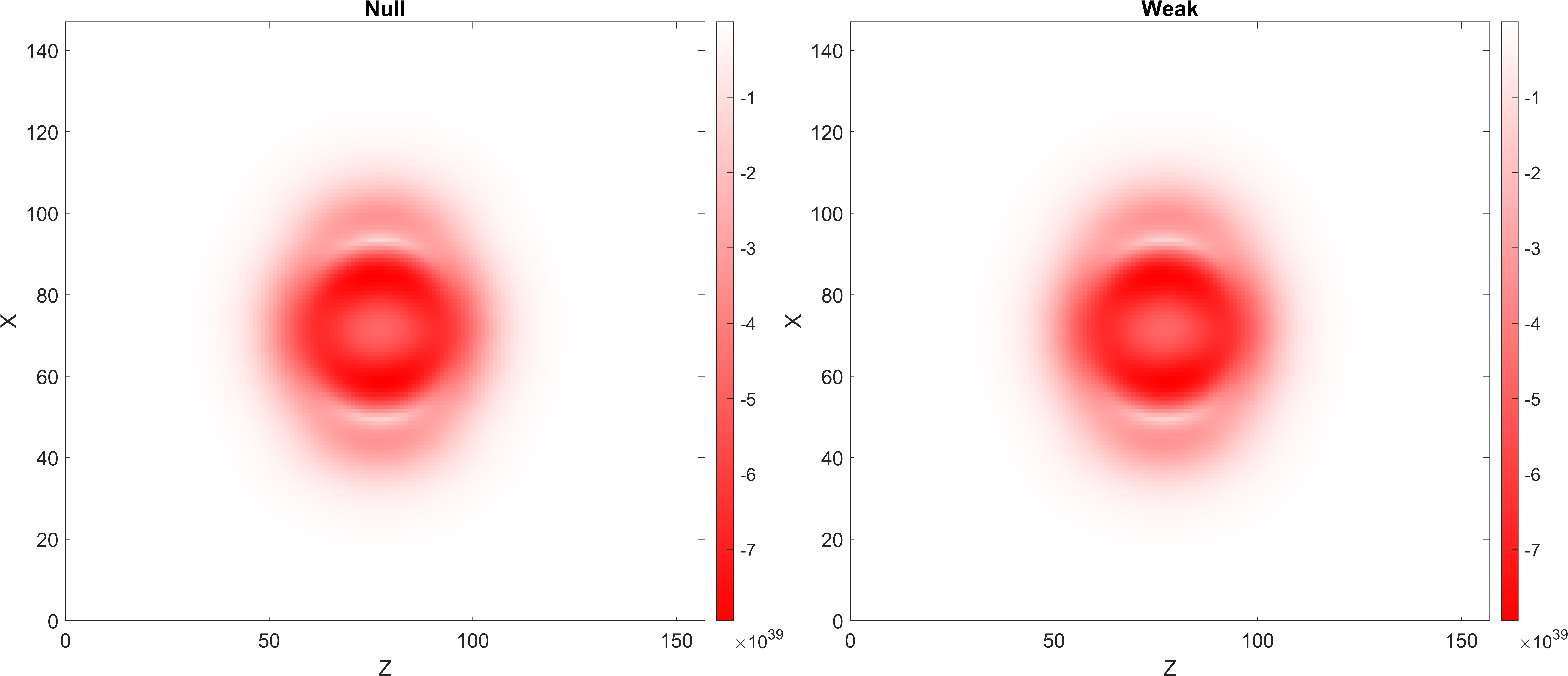}
\caption{NEC and WEC for the starting metric. Direction of motion is along +Z. Cross-section is plotted for y = 0.}\label{fig:Opt_EcondStart}
\end{figure}

\begin{figure}[hbt!]
\centering
\includegraphics[width=\textwidth]{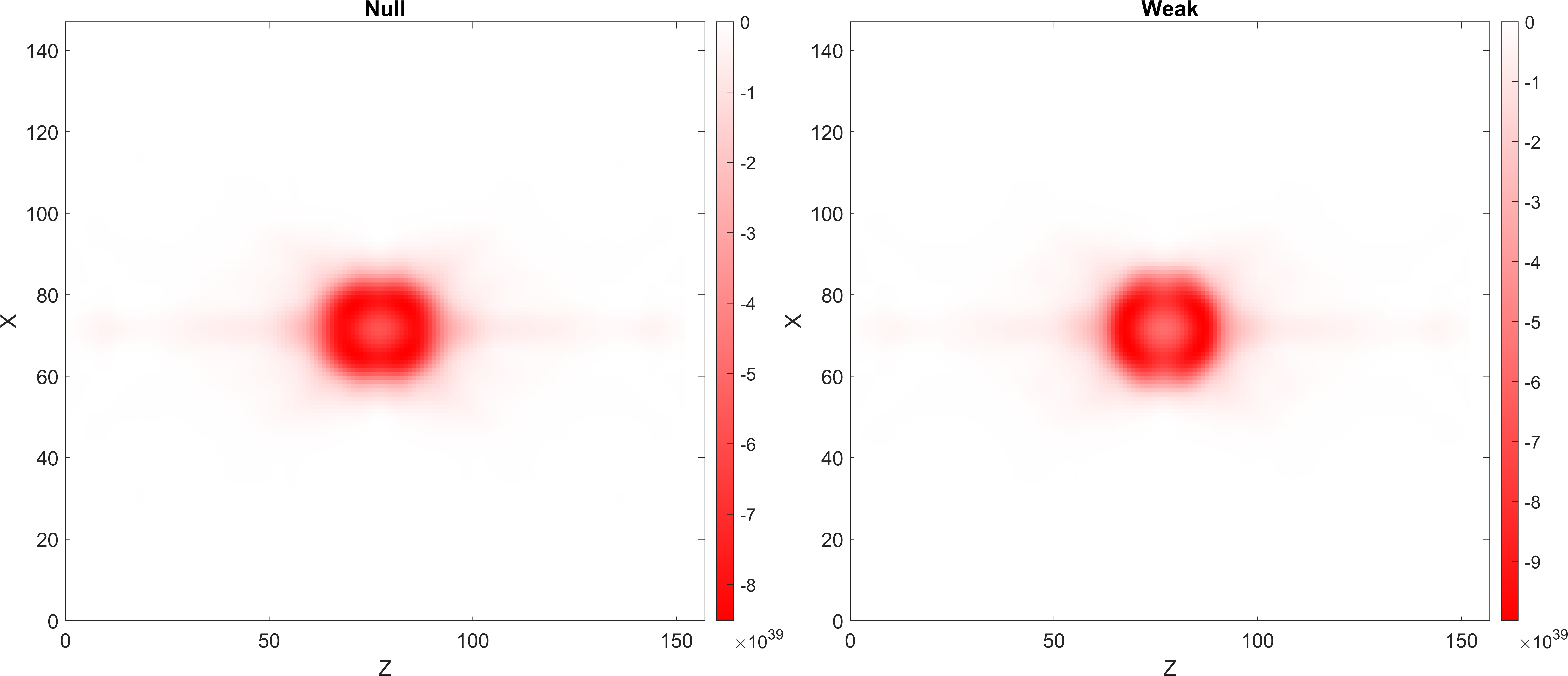}
\caption{NEC and WEC for the optimized metric. Direction of motion is along +Z. Cross-section is plotted for y = 0.}\label{fig:Opt_EcondEnd}
\end{figure}

\begin{table}[h]
    \centering
    \caption{Optimization parameters run on the starting AM.}
    \begin{tabular}{|p{3cm}|c|p{9cm}|}
        
        \multicolumn{1}{p{3cm}}{Parameter} & \multicolumn{1}{c}{Value} & \multicolumn{1}{p{6cm}}{Description} \\
        \hline
        \textit{Passes} & 10  & Number of passes that a given perturbation is allowed to add if an improvement is made.\\
        \hline
        \textit{Generations} & 540  & Number of optimization runs over a population. \\
         \hline
         \textit{Population} & 18 & Number of parallel optimizations run during a given generation.\\
         \hline
         \textit{Perturbation Terms (6+4)} & $\alpha, \beta_2, \beta_3, \gamma_{2}, \gamma_3, \kappa_{2}$ & Allowed metric functions to vary.\\
         \hline
    \end{tabular}
        \label{tab:OptStartData}
\end{table}
The fitness metric used is the NEC, which takes all values of $\nu_{NEC} < 0$ and integrates them across the entire volume. The starting metric has a fitness value of $7.2 \times 10^{44}$\footnote{This is negative in nature, but the absolute value is used, with any reduction of this value being the improvement factor.}. Following the optimization run, the fitness is improved by a factor 3x to $2.1 \times 10^{44}$. Since the optimization can vary lapse, shift vectors, expansion, and shear terms, the final metric generated differs significantly from the starting AM. The resulting metric functions are shown in Figure \ref{fig:Opt_Functions}.
The Eulerian energy density for the starting and optimized metric is shown in Figure \ref{fig:Opt_EdenStart} and \ref{fig:Opt_EdenEnd} respectively. The key difference is a contraction of the bubble energy density which is also accompanied by the addition of positive energy regions beyond it. A similar impact is seen in the energy condition violations as they are also pushed into the center region, shown in Figures \ref{fig:Opt_EcondStart} and \ref{fig:Opt_EcondEnd}.

\begin{figure}[hbt!]
\centering
\includegraphics[width=\textwidth]{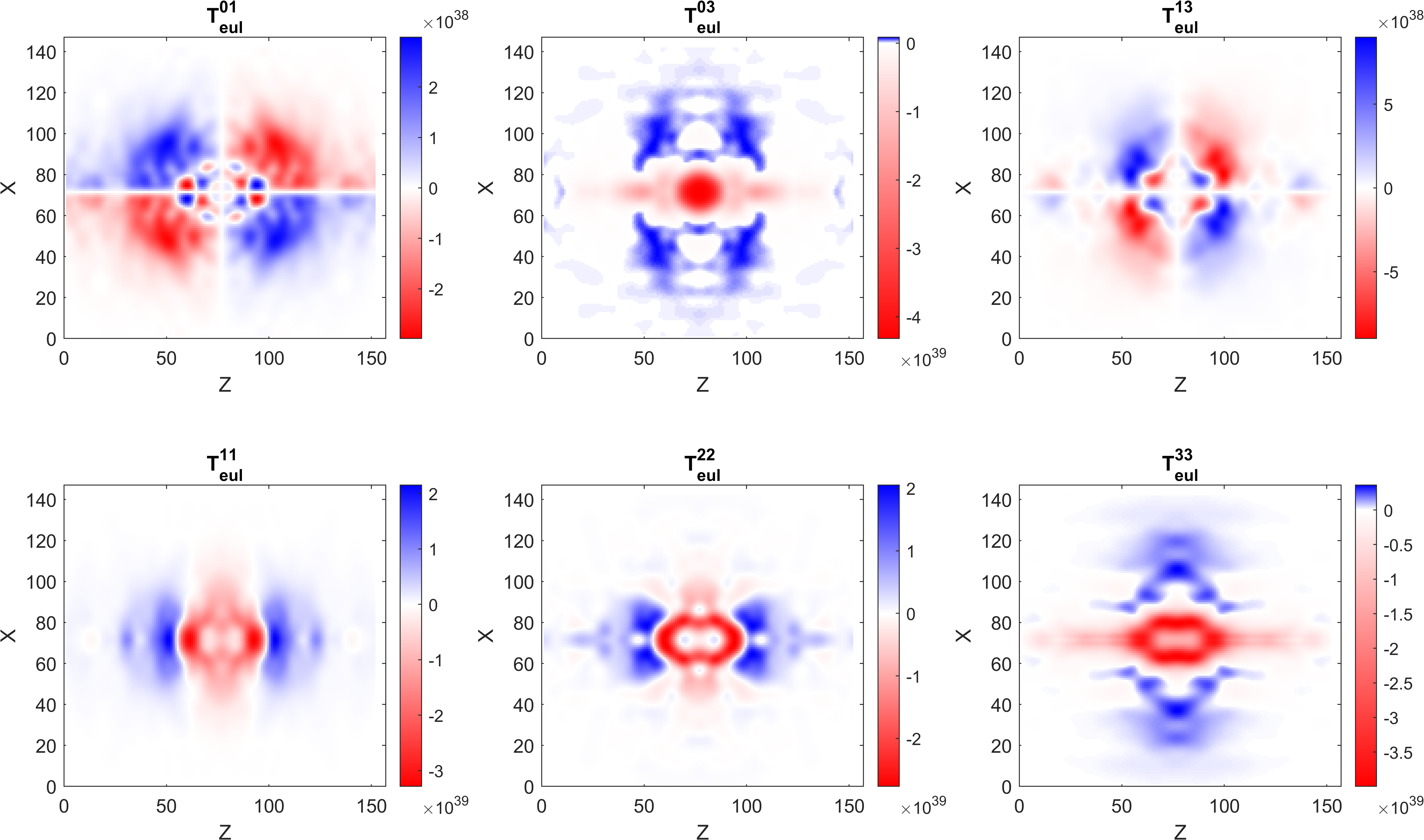}
\caption{Eulerian stress-energy tensor components for the optimized metric. The $T^{00}$ component is shown in Figure \ref{fig:Opt_EdenEnd}. Direction of motion is along +Z. Cross-section is plotted for y = 0.}\label{fig:Opt_Etensor}
\end{figure}

Looking at the full stress-energy components, shown in Figure \ref{fig:Opt_Etensor}, a much more intricate structure emerges with complex momentum flows and pressure regions. This complexity is partly due to the nature of the Gaussian perturbations themselves, which tend to cause "noisy" patches. A Gaussian function can approximate any kind of shape, but with a Monte Carlo approach, it can take a very long time to smooth out the wrinkles toward the final optimum solution, which itself is hard to determine in such a complex phase space. 

\begin{figure}[hbt!]
\centering
\includegraphics[width=\textwidth]{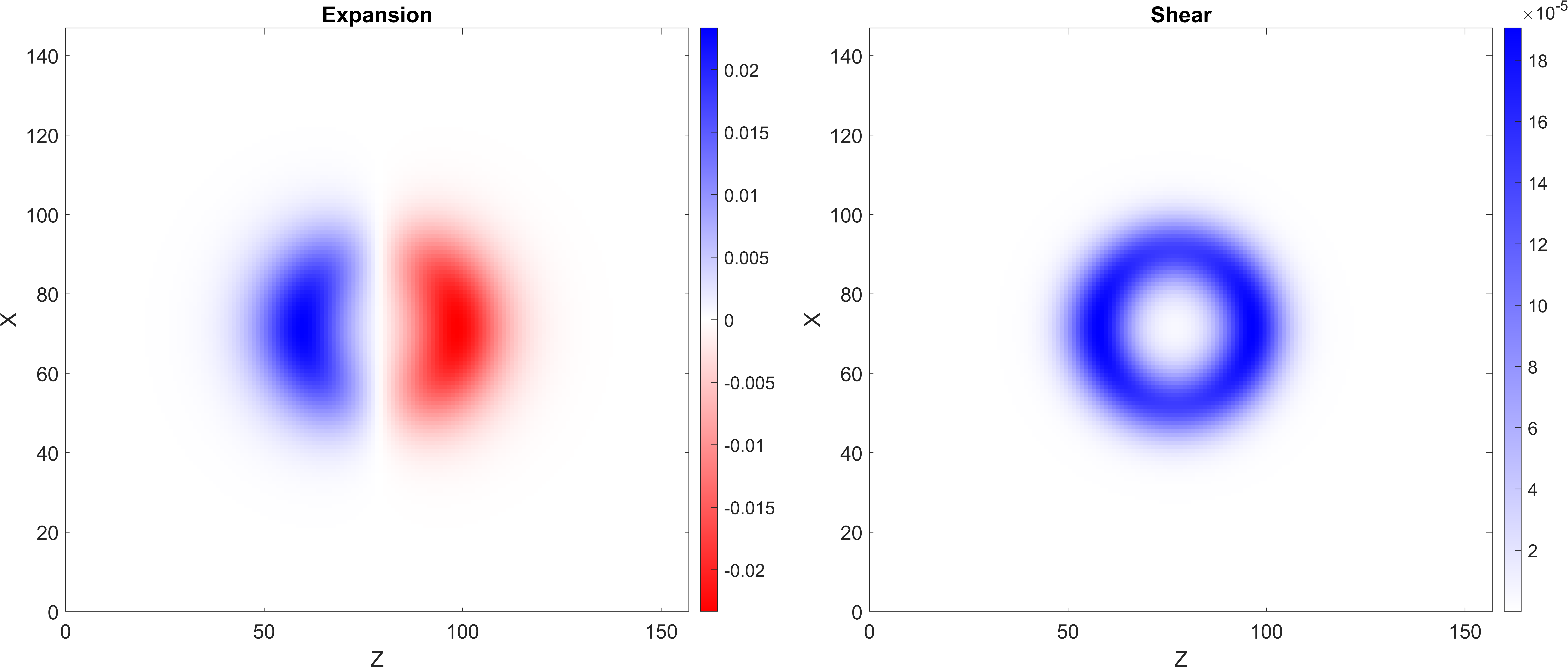}
\caption{Expansion and shear scalars for the starting metric. Direction of motion is along +Z. Cross-section is plotted for y = 0.}\label{fig:Opt_scalarsStart}

\end{figure}

\begin{figure}[hbt!]
\centering
\includegraphics[width=\textwidth]{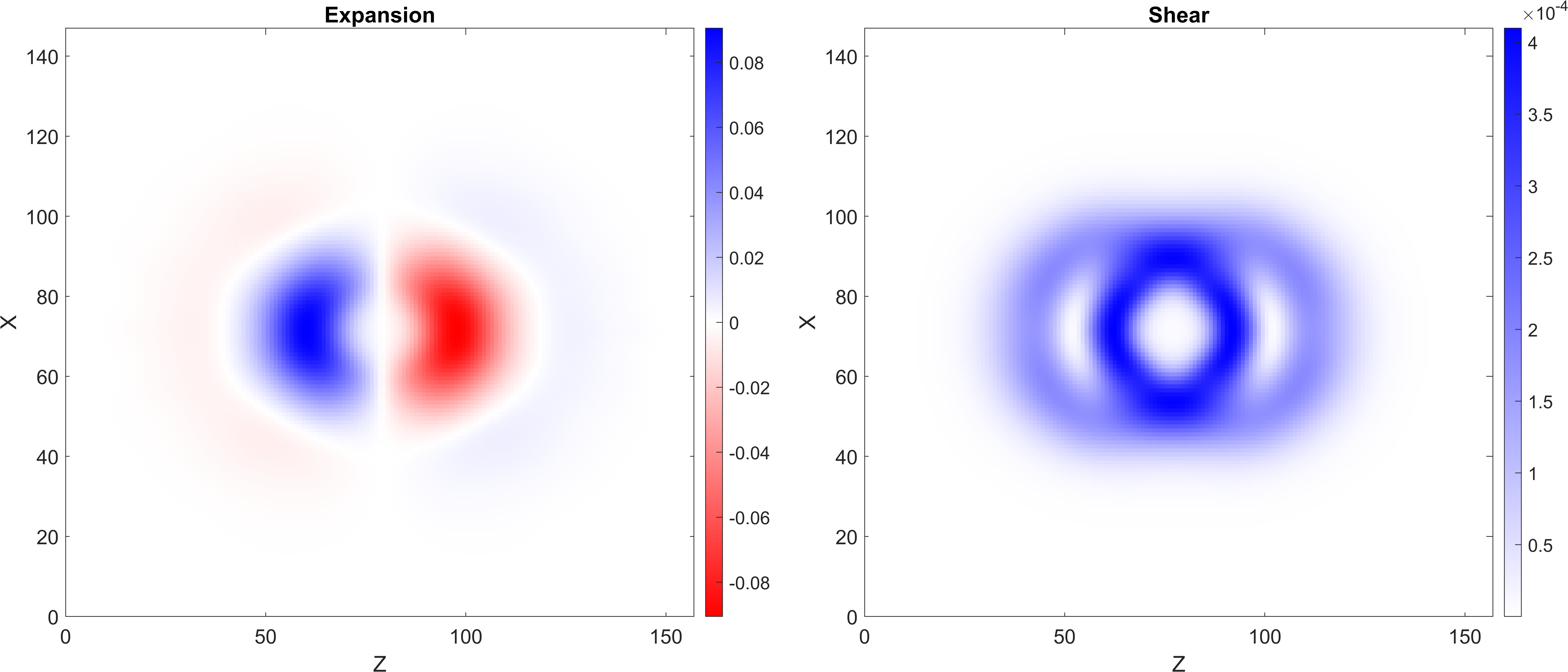}
\caption{Expansion and shear scalars for the optimized metric. Direction of motion is along +Z. Cross-section is plotted for y = 0.}\label{fig:Opt_scalarsEnd}
\end{figure}

Finally, comparing the scalars in Figures \ref{fig:Opt_scalarsStart} and \ref{fig:Opt_scalarsEnd}, the optimization resulted in the addition of concentric regions of the expansion and contraction. The inner region maintains the general expansion behind and contraction in front being slightly flattened along +X, but an outer region emerges which is opposite in sign. Again, it should be noted that the existence of expansion and contraction is not a requirement for a warp drive but can exist depending on the nature of the metric \cite{2002CQGra..19.1157N}. Also differing from the AM is the shear, which again is everywhere positive but sees an addition of an inner and outer ring.


The optimization performed here is only an example case to demonstrate its function. The method can be extended through the use of different fitness functions. For example, the NEC is useful in removing negative energy density, but often it results in the addition of positive energy to the world. Furthermore, the world and passenger volume boundaries restrict the possibility to remove global violations. If the defined world inherently has violations at the boundary (towards infinity), then the optimization is unable to change the root cause of the violation, often adding undesirable changes, such as additions of positive matter up to the world boundary. In general, the optimizer is best at removing local violations or optimizing the energy needed. Care must be taken to ensure that trivial solutions are not being generated. Certainly, a metric that does solve these issues at the onset can likely be optimized for its required energy, which will be an important step in addressing the reasonable mass requirement of a practical warp drive.

\clearpage

\section{Conclusion}\label{sec:Conclusion}
Development of the Warp Factory codebase offers a powerful tool for the development of warp drives. The inherent challenges of solving the EFE present a hurdle to exploring novel warp metrics. The primary issue in developing warp metrics is overcoming the problems of its physicality, which requires a complicated analysis of timelike observers on the energy tensor components. To date, most analytical work considers a limited case of observers, which, for complicated metrics, might hide certain violations. The physicality of warp drives is not just a requirement of positive energy density, it is also the relation between momentum density, pressures, and shears within the stress-energy tensor. Evaluating this in a numerical framework allows users to establish and understand the dynamics of various warp metrics in a fully general manner. Warp Factory provides a tool for the visualization and analysis of arbitrary metrics with a focus on the evaluation of energy conditions, which is a unique capability when compared to most numerical GR tools.

In this paper, the functionality of Warp Factory has been demonstrated through the evaluation of classic and recent warp metrics, providing unique 3D analysis and visualizations of their stress-energy tensors, energy conditions, and scalars. One important insight is that the Lentz metric, which has been proposed as possibly solving the WEC violation, does still violate the WEC when analyzed across the space of timelike observers. Investigations into the impact of different passenger volumes on energy requirements and energy condition violations found that the overall shape has a marginal impact.


Warp Factory also implements an optimization module that provides a unique approach to warp research. While the method is computationally intensive and prone to limitations due to the large parameter space being explored, it offers insight into what elements of the warp metric tend to improve the energy conditions. It also allows changes to all metric functions as decided by the user, creating complicated structures not easily found or expressed analytically. The initial application of this to the AM sees some improvement in reducing violations through unique changes to the metric functions.

The APL team's next steps for Warp Factory are to expand the energy condition evaluation to include dominant and strong conditions while preparing the codebase for public use on Github. In addition, the team is developing functions for geodesic solvers to compute light rays and matter trajectories around a warp drive in a general 3D case. The team anticipates publishing follow-up papers on the results of these efforts.


\section*{Acknowledgments}
Special thanks to the entire APL team and Sarah Dangelo for review and comments.

\bibliography{papermain.bib}

\end{document}